\documentclass[twocolumn, superscriptaddress, secnumarabic, amssymb, nobibnotes, nofootinbib, aps, prd]{revtex4}
\usepackage{amsmath}
\usepackage{amsfonts}
\usepackage{amssymb}
\usepackage{comment}
\usepackage{float}
\usepackage{caption}
\usepackage{subcaption}
\usepackage{graphicx}
\usepackage{appendix}
\usepackage{color}
\usepackage{tikz}
\usepackage{soul}
\usepackage{ulem}{%for delete line%}
\usetikzlibrary{quotes, angles}
\usepackage[unicode=true,
bookmarks=false,
breaklinks=false,pdfborder={0 0 1},backref=false,colorlinks=true]
{hyperref}
\hypersetup{
	urlcolor=green,linkcolor=blue,citecolor=red}

\usepackage{soul}
\setstcolor{blue}

\begin{document}
	\title{Phonon-induced contrast in a matter-wave interferometer}
	
	\author{Qian Xiang}
	\affiliation{Van Swinderen Institute, University of Groningen, 9747 AG, The Netherlands}

	\author{Run Zhou}
	\affiliation{Van Swinderen Institute, University of Groningen, 9747 AG, The Netherlands}

	\author{Sougato Bose}
	\affiliation{Department of Physics and Astronomy, University College London, Gower Street, WC1E 6BT London, UK}
	
	\author{Anupam Mazumdar}
	\affiliation{Van Swinderen Institute, University of Groningen, 9747 AG, The Netherlands}

	\begin{abstract}
		Utilizing the Stern-Gerlach apparatus to create matter-wave superposition states is a long-sought-after goal, not only due to its potential applications in the quantum realm but also because of its fundamental implications for studying the quantum properties of gravity. The main challenge in creating a macroscopic quantum interferometer arises from the loss of coherence, primarily through two channels. One channel involves strong coupling with the environment for macroscopic matter, leading to decoherence. The other channel relates to the precision of wave packet overlap, which can occur due to external and internal fluctuations of various sources. The latter introduces a unique challenge for larger-scale masses by perturbing the centre of mass motion of the macroscopic object. Here, we study a particular challenge, namely, the issue of internal degrees of freedom, specifically phonon fluctuations and contrast reduction. This work will investigate the contrast reduction caused by spin-magnetic field-phonon and diamagnetism-phonon interactions in the quantum gravity-induced entanglement of masses (QGEM) protocol configuration.
		
	\end{abstract}
	
	\maketitle
	\section{Introduction}\label{Introduction}

	The matter-wave interferometer is an essential tool for generating sizeable spatial superposition states. It represents the forefront of theoretical physics, as the creation of spatial superposition states can aid in constructing a theory that reconciles gravity with quantum mechanics\cite{Penrose:1996cv,PhysRevA.40.1165,Pearle:1988uh,Bassi:2012bg,PhysRevLett.110.160403}, studying the equivalence principle\cite{
PhysRevLett.120.183604,PhysRevLett.125.191101,Bose:2022czr,Chakraborty:2023kel}, decoherence mechanism\cite{
 ORI11_GM,PhysRevLett.107.020405,Romero-Isart:2009uui,PhysRevA.104.052416,PhysRevA.105.032411, vandeKamp:2020rqh,Rijavec:2020qxd, Gunnink:2022ner,
Schut:2021svd, Schut:2023tce, Fragolino:2023agd,Toros:2020krn,Schut:2023eux}. In terms of applications, matter-wave interferometers are also considered for detecting gravitational waves\cite{Marshman:2018upe}, neutrinos~\cite{PhysRevResearch.5.023012}, detecting light axion-like particles~\cite{Barker:2022mdz}, for precise measurements of Earth's gravitational acceleration\cite{Peters:1999tya,Chiao:2003sa,PhysRevD.73.084018,PhysRevD.73.022001,PhysRevD.78.122002,DIMOPOULOS200937,Tino:2019tkb,PhysRevLett.110.171102, Marshman:2018upe}, and detection of space debris~\cite{Wu:2022rdv}.
	
	To test the quantum nature of gravity, a proposal known as quantum gravity induced entanglement of masses (QGEM) has been recently proposed~\cite{Bose:2017nin,marletto2017gravitationally}. To further probe the nature of gravity, the gravitational-optomechanical test, which tests the analogue of light-bending experiment in quantum gravity~\cite{Biswas:2022qto}, and measurement-based tests~\cite{Hanif:2023fto}, all require the creation of a matter-wave interferometer.
	
	The QGEM proposal utilizes the two masses, each prepared in a spatial superposition state, which is kept at a distance where the only interaction between them is mediated via gravity. Suppose detectable entanglement exists between these two masses. In that case, one can ascertain that the gravitational interaction is quantum in nature~\cite{Marshman:2019sne,Bose:2022uxe,Vinckers:2023grv,Elahi:2023ozf,Christodoulou:2022vte,Carney_2019,Danielson:2021egj}. This quantum-entanglement-based, experimental is concordant with the local operations and classical communication (LOCC) theorem, which states that classically one cannot mediate quantum entanglement between the two quantum systems~\cite{Bennett_1996}. 
	
	In QGEM, the spatial superposition is created by the Stern-Gerlach (SG) apparatus \cite{book1}. At the same time, the test mass is played by a diamond internally embedded with a nitrogen-vacancy centre (NV). The SG device utilizes the coupling between the magnetic field and the spin (NV-centre) to create a spatial superposition state. Based on similar mechanisms, there have been approaches toward atom interferometry. Such interferometer has been experimentally realized on atom chips for  half-loop~\cite{machluf2013coherent} and full-loop \cite{Margalit:2020qcy} schemes. For the former, the coherent time for maintaining spatial superposition was $21.45$ ms, and the size of 3.98 $\mu$m was reached. For the full loop, the coherent time and superposition size were 7 ms and 0.38 $\mu$m, respectively. 
	
	Of course, QGEM imposes higher demands on the interferometer's performance, with critical requirements including the diamond's mass $M \sim 10^{-15}- 10^{-14}$ kg (with length scale $\sim 10^{-7}- 10^{-6}$ m) and the spatial superposition size needing to reach approximately $10-100 \mu$m~\cite{Bose:2017nin,vandeKamp:2020rqh,Schut:2023eux,PhysRevA.105.032411}, and \cite{Schut:2023hsy}. The larger mass objects tend to couple more strongly with their environment, such as the electromagnetic fields and collisions with residual gas particles. 
	These couplings act as measurements, causing the collapse of the spatial superposition state into classical states, known as decoherence~\cite{joos2003decoherence,Schlosshauer:2019ewh}. 
	Furthermore, due to experimental constraints on the magnetic field gradient~\cite{Marshman:2021wyk,Marshman:2023nkh}, the time required to achieve large-scale spatial superposition states is on the order of $\Delta t \sim 1$ second makes it more challenging to maintain the spin coherence state of NV centres and the spatial superposition state of diamond~\footnote{See also gravity gradient and relative acceleration noise~\cite{Toros:2020dbf,Grossardt:2020,
			Wu:2022rdv} for other sources of noise that lead to dephasing the interferometer along with electromagnetic sources of noise~\cite{Schut:2023tce,Fragolino:2023agd}.}.
   
	For a closed-loop scheme, another challenge that is independent of the environment arises from the Humpty-Dumpty (HD) effect\cite{Englert1988-ENGISC,schwinger1988spin,scully1989spin,englert1997time}, which refers to the overlap problem of matter waves. For example, in the atomic interferometer mentioned above, if the wave packets on both sides of the interferometer arm cannot precisely overlap (both in translational and rotational degrees of freedom), they cannot generate interference fringes and thus lead to the loss of contrast, see~\cite{Japha:2022xyg} where the analysis of HD effect has been analysed quite nicely including the impact of angular momentum and the libration mode of the NV spin.
	For macroscopic objects, there is also a challenge of overlap due to the internal degrees of freedom, such as phonon vibration. 
	In a very interesting work\cite{Henkel1}, the contrast reduction of phonon wave packets induced by coupling NV centres with magnetic fields was demonstrated. More recently, this coupling has been extended to white noise, providing a boundary on creating larger-scale spatial superposition states that have been considered~\cite{Henkel2}.

	This work will also investigate the contrast loss caused by phonons in matter-wave interferometers.
	Unlike atomic-scale interferometers, due to the diamagnetic property of a diamond and its interaction with the magnetic field being proportional to its mass, the diamagnetic interaction becomes essential in the matter-wave interferometers with a massive object. For example, the diamagnetic interaction contributes to the trajectories of the diamond's centre of mass (CoM) wave packets. Therefore, the magnetic field must be carefully designed to guarantee a close loop\cite{Marshman:2021wyk,Zhou:2022epb,Zhou:2022frl,Zhou:2022jug}. Besides, there has been recently proposed a mass-independent scheme\cite{Zhou:2022jug} based on diamagnetic interaction of creating large superposition size. From the perspective of internal degrees of freedom, the diamagnetic repulsion of the diamonds on either side of the interferometer arms are different, leading to the excitation of varying phonon modes and thus resulting in a contrast reduction of internal degrees of freedom. This is fundamentally similar to the spin-magnetic field interaction papers~\cite{Henkel1,Henkel2}. 
	
\begin{figure*}
		\begin{subfigure}[]{0.45\textwidth}
			\includegraphics[width=\textwidth]{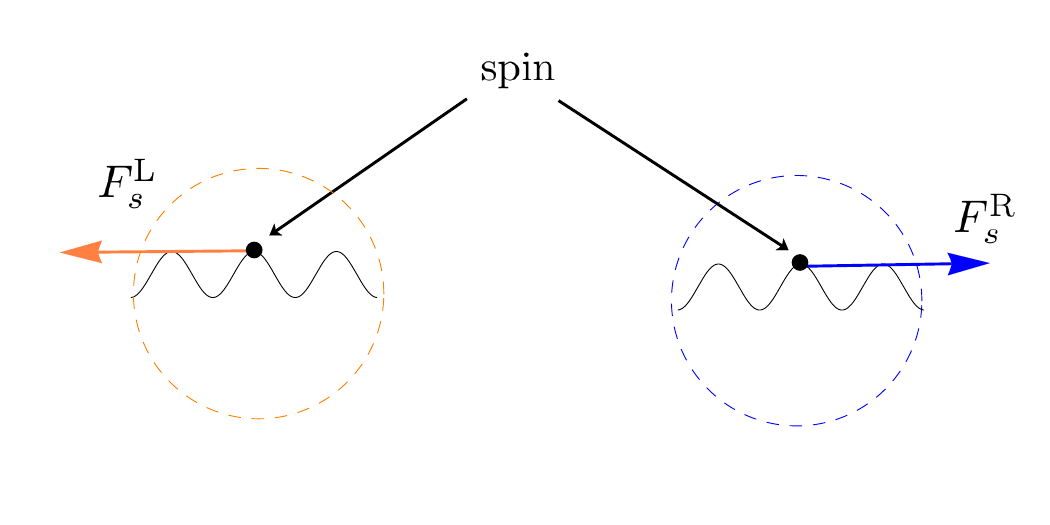}
			\subcaption*{(a)}
			\label{Fig1Spin}
		\end{subfigure}
		\begin{subfigure}[]{0.45\textwidth}
			\includegraphics[width=\textwidth]{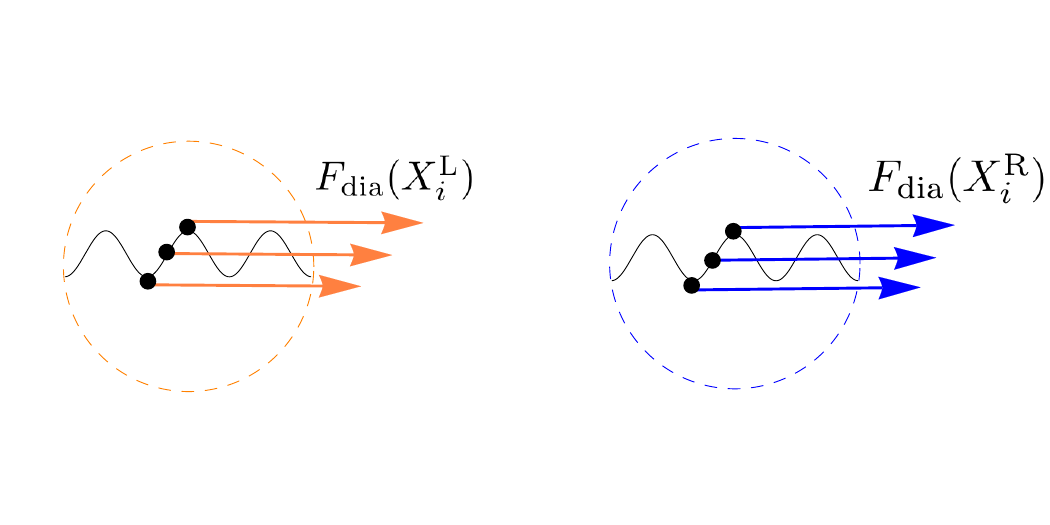}
			\subcaption*{(b)}
			\label{Fig1Dia}
		\end{subfigure}
		\caption{
			\small {Illustrations of different kinds of forces acting on the lattice chain (represented by black solid wave lines) where positions of atom/spin are dotted in black. The wave packets of the centre of mass (CoMs) in the two interferometry arms denoted here by the left (L) and right (R) arms are plotted by orange and blue circles. Panel (a) shows splitting forces $F_s^\text{L,R}$ acting on the spin with site number $s$ where they have the same magnitude but opposite direction. Panel (b) shows diamagnetic forces $F_i^{\text{L,R}}$ acting on every atom, where, for simplicity, we assume it acts on the spin as well}. }
		\label{figure1}
	\end{figure*}

	In this regard, our paper differs from the Refs.\cite{Henkel1,Henkel2}. We validate their results for the spin-magnetic field interaction. However, we also 
	consider the effect of diamagnetic contribution, which brings another layer of complexity that is not present otherwise. In the case of diamagnetic contribution, the phonons are also coupled to the CoM of the diamond. The difference lies in the fact that for diamagnetic interaction, each carbon atom in the atomic chain on both sides of the interferometer arms experiences repulsion in the same direction, while for the spin-magnetic interaction, the chains are plucked at the NV-centre in opposite directions, as illustrated in Fig. \ref{figure1}. Besides the magnetic effects, we will also briefly mention the role of electric dipole interaction with phonons and how it affects the latter in comparing the loss of contrast for the one-loop interferometer.
	
	The structure of this work is arranged as follows. In section \ref{Set Up}, we introduced the setup, which includes the phonon model and the statistical model of phonon modes at finite temperature, and introduced the Wigner function. In \ref{Interactions in SGI}, we provided the Hamiltonian of the diamond and presented the model for the trajectory of the diamond wavepacket. 
	In \ref{secPhonon}, we examined the different effects of diamagnetism-phonon coupling and spin-phonon interaction on phonon dynamics and further utilized displacement operators in Section \ref{Loss of Contrast} to describe the phonon's contrast reduction. In Section \ref{Results}, we presented numerical results and compared the contrast loss induced by these two interactions at different temperatures. Also, we have considered contrast loss led by other interactions, potential dipole-electric field interaction, in the level of phonon; this part of the discussion can be found in Appendix \ref{dipole1} and \ref{dipole2}.

	%%%%%%%%%%%%%%%%%%%%%%%%%%%%%%%%%%%%%%%%%%

	%%%%%%%%%%%%%%%%%%%%%%%%%%%%%%%%%%%%%%%%%%%%%%%%%%%%%%%%%%%
	\section{Brief review on Phonon }\label{Set Up}
	Phonons constitute the primary focus of this work, and the model is derived from the lattice chain model, see Ref.\cite{jishi2013feynman}. In the case of atomic interactions, the Hamiltonian for free phonons can be expressed as follows,
	\begin{equation}
		\begin{aligned}
			H_\text{free}&= m \sum_i \frac{\dot{x}_i^2}{2}  + \sum_{i,j} \frac{K_{ij}}{2}x_i x_j\\
			&= \sum_q \frac{1}{2}\left(\dot{u}_q^2+\omega_q^2 u_q^2\right)\,,
		\end{aligned}\label{Free}
	\end{equation}
	where $K_{ij}$ is a matrix associated with the inter-atomic interaction while $Q_q(X_i) = Q_q^i$ represents the eigenvector of the matrix. The position of each atom can be written as $X_i(t) + x_i(t)$, where the capital $X_i (t)$ denotes the equilibrium position of the $i^\text{th}$ atom at any instant of time, and $x_i$ represents its small vibrations. Because the position of the diamond as a whole varies within the SG apparatus, the equilibrium position $X_i(t)$ of each atom in the chain becomes a function of time. Notably, the relative distances between these equilibrium positions, $X_{i+1}-X_i= \text{const.}$, remain constant. Therefore, $X_i(t)$ can effectively be regarded as the trajectory of diamond's CoM, with their distances relative to the centre particle $X_\text{c}$ remaining unchanged. Here, for convenience, we assume that the NV centre (with equilibrium position $X_\text{s}$) is embedded at the CoM position, $X_\text{s}(t)=X_\text{c}(t)$. In the following, we sometimes neglect the time $t$ for convenience.

	The small movement $x_i(t)$ can be expanded into mode amplitude $u_q$, with creation and annihilation operators as,  
	\begin{equation}
		\begin{aligned}
			x_i(t)&=\frac{1}{\sqrt{Nm}}\sum_q  Q_q^i u_q(t),\\
		\end{aligned}\label{Free1}
	\end{equation}
	where the relation between $u_q$ and its conjugate (the momentum $\dot{u}_q$), and the creation and annihilation operator can be expressed as,
	\begin{equation}
		\begin{aligned}
			a_q=\frac{1}{\sqrt{2\hbar \omega_q}}(\omega_q u_q+\text{i}\dot{u}_q),\\
			a_{-q}^\dagger=\frac{1}{\sqrt{2\hbar \omega_q}}(\omega_q u_q-\text{i}\dot{u}_q).
		\end{aligned}\label{Free2}
	\end{equation}
	For simplicity, in our one-dimensional atomic chain model embedded with an NV centre, we assume that all $N$ particles have an identical atomic mass of carbon $m$ and the mass for the diamond is $M$. The $\omega_q$ is the eigenfrequency of the chain model with the fundamental tone,
 \begin{equation}
		\begin{aligned}
			\omega_0={\pi c}/{L}
		\end{aligned}\label{tone}
	\end{equation}
 where $c=17.5\times 10^{3} {\rm m/s}$ is the sound speed within a diamond~\cite{jishi2013feynman} and $L$ is the length (estimated via $L= (M/\rho)^{-\frac{1}{3}}$, where the density $\rho=3.51\times 10^{3}$ kg/m$^3$ \cite{greenwood2012chemistry}). For example, for the masses in a range $M \sim 10^{-14}- 10^{-18}$~kg the corresponding fundamental tone ranges from $\omega_0 \sim 10^{10}- 10^{11}$~Hz.
	
	At the initial stage, $t=0$, before the diamond enters the SG apparatus and interacts with the magnetic field, one can assume that the phonons are in a thermal equilibrium state at a temperature $T$~\cite{jishi2013feynman},
	\begin{equation}
		\begin{aligned}
			\langle n_q(t=0)\rangle=\frac{1}{2}\coth\left(\frac{1}{2}\frac{\hbar \omega_q}{k_\text{B} T}\right).
		\end{aligned}\label{nq0}
	\end{equation}
	According to the equipartition theorem, one can establish the relationship between the average number of phonons $\left\langle n_q(0)\right\rangle$, the characteristic length $\sigma_{u_q}$ and the characteristic momentum $\sigma_{\dot{u}_q}$ as~\cite{jishi2013feynman},
	\begin{equation}
		\begin{aligned}
			\langle n_q(t=0)\rangle=\frac{\omega_q }{\hbar}\sigma^2_{u_q}=\frac{1}{\hbar \omega_q}\sigma^2_{\dot{u}_q},
		\end{aligned}\label{12}
	\end{equation}
	where $\sigma_{u_q}$ and $\sigma_{\dot{u}_q}$ are statistical averages. A review of this detail can be found in the appendix \ref{appen2}. 
	In the case of thermal equilibrium, the phonons in each mode $q$ form a mixed state. The amplitudes and momenta of the phonons thus take on Gaussian distributions, represented by the Wigner function\cite{Pedestrain} as follows,
	\begin{equation}
		W\left(u_q, \dot{u}_q\right)={\cal{N}} \exp \left[-\left(\frac{u_q^2}{2 \sigma_{u_q}^2}+\frac{\dot{u}_q^2}{2 \sigma^2_{\dot{u}_q}}\right)\right],
		\label{Wigner}
	\end{equation}
	where $\cal{N}$ is normalization factor. Eq.~(\ref{12}) gives the normal mode's characteristic length and momentum. Note that $\sigma_{u_q}$ and $\sigma_{\dot{u}_q}$ are statistical averages dependent on temperature $T$, see appendix \ref{appen2}. Once we let the diamond experience the SG force, the occupation number of the phonons will evolve, and we will estimate that in the following sections.
	
	%%%%%%%%%%%%%%%%%%%%%%%%%%%%%%%%%%%%%%%%%%%%%%%%%%%%%%%%%%%%%%%%%%%%%%%%%%%%

	\section{Interactions in Stern-Gerlach Apparatus}\label{Interactions in SGI}	
	
	For the diamond embedded with a single spin $S$ in the NV, we already assume that it has been prepared in a superposition of opposite direction $S^\text{R}=1$ (right) and $S^\text{L}=-1$ (left), hence the spatial superposition. The Hamiltonian for the diamond's CoM consists of three parts~\footnote{There is also a term related to $D(S\cdot \hat n)^2$ related to the zero-point splitting~\cite{Japha:2022xyg,Zhou:2022epb}. This term is relevant for the internal spin degrees of freedom. Here, we are not considering this contribution; we do not consider the rotation of the diamond in our current context. 
    For the rotation of NV centre in diamond and its mechanical coupling with magnetic levitation, see \cite{delord2020spin,perdriat2021spin}.},
	\begin{equation}
		\begin{aligned}
			H_\text{CoM}= \frac{P_c^2}{2M}-\mu S^\text{L,R} B-\frac{\chi_{\rho} M}{2\mu_0}B^2,
		\end{aligned}\label{CoM}
	\end{equation}
	the first term is just the kinetic energy relating to CoMs's momentum. The second term is the the coupling between the NV centre and the magnetic field with the magnetic moment $\mu=g \mu_\text{b}$, where the Land\'e g-factor $g\approx2$ and Bohr magneton $\mu_\text{b}=9.27\times 10^{-24}$ (J/T) \cite{o1999modern}. For simplicity, we assume the NV is located at the CoM with equilibrium position $X_\text{s}=X_\text{c}$. The appearance of the third term is because diamond is a diamagnetic material, and its susceptibility is a negative value $\chi_\rho = -6.2 \times 10^{-9} {\rm m^3/kg}$ \cite{PhysRevB.49.15122} with vacuum permeability $\mu_0$. One should note that this term is a macroscopic expression. When a carbon atom chain is placed in an external magnetic field pointing in the $\mathbf{e}_x$ direction, each carbon atom will contribute a different magnitude of repulsion according to its equilibrium position $X_i$ pointing in the $-\mathbf{e}_x$ direction. From the aspect of  macroscopic diamond, these small repulsion collectively form a macroscopic diamagnetic force effectively acting on the CoM, $X_\text{c}$.
	
	For the magnetic field in Eq.~(\ref{CoM}), we give its form that is experienced by the $i^\text{th}$ atom as,
	\begin{equation}
		\begin{aligned}
			B_i(t)= B_0+ b(t) X_i(t),
		\end{aligned}\label{BField}
	\end{equation}
	where $B_0$ is the bias magnetic field, and $b(t)$ denotes the gradient in the SG.
	Note that in Eq.(\ref{BField}), the coordinates along the $\mathbf{e}_x$ axis are represented by the uppercase $X_i$, indicating the equilibrium positions of atoms in the chain. For example, the magnetic field experienced by the spin, which we have assumed to be located at the centre of the chain, becomes $B_\text{c}=B_0+b(t) X_\text{c}$. In this model, we only consider the one-dimensional case, where the separation of the diamond and the force on the atomic chain is confined to the $\mathbf{e}_x$ direction. 
	
\begin{figure*}[t!]
 \begin{subfigure}[]{0.32\textwidth}
			\includegraphics[width=\textwidth]{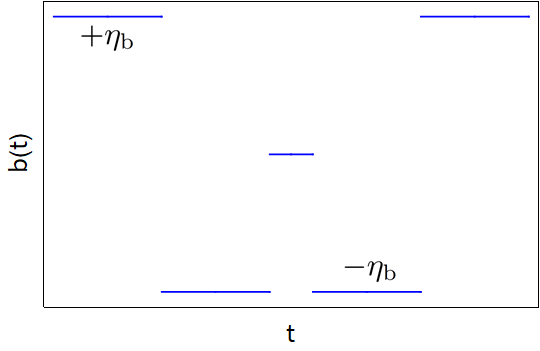}
			\subcaption*{(a)}
		\end{subfigure}
		\begin{subfigure}[]{0.32\textwidth}
			\includegraphics[width=\textwidth]{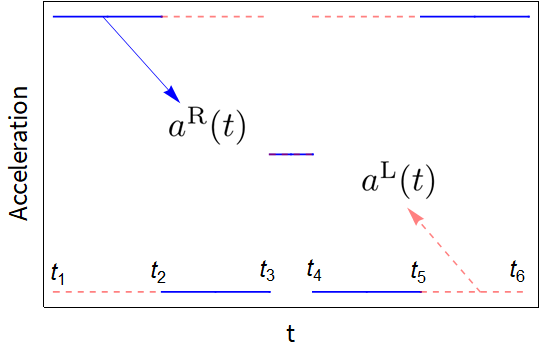}
			\subcaption*{(b)}
		\end{subfigure}\\
		\begin{subfigure}[]{0.32\textwidth}
			\includegraphics[width=\textwidth]{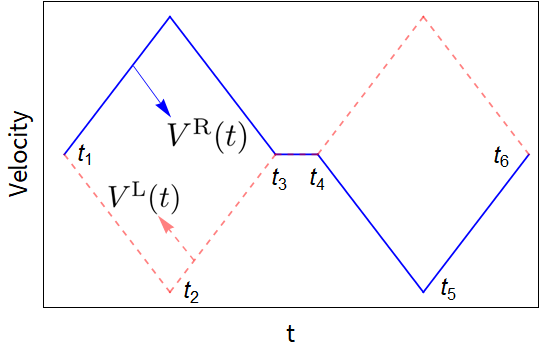}
			\subcaption*{(c)}
		\end{subfigure}
		\begin{subfigure}[]{0.32\textwidth}
			\includegraphics[width=\textwidth]{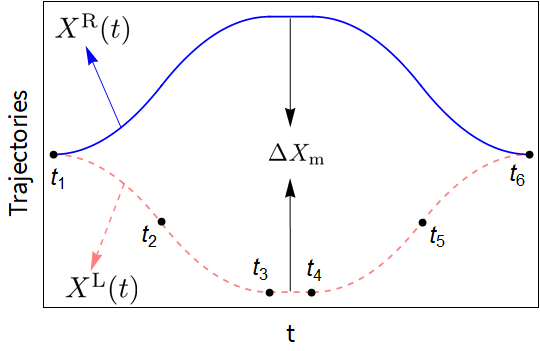}
			\subcaption*{(d)}
		\end{subfigure}
		\caption{
			\small {The four sub-figures illustrate our magnetic field setting (a) inside the Stern-Gerlach apparatus and the corresponding accelerations (b), velocities (c) and trajectories (d) of CoM wave packets. In the time intervals $(t_1,~t_2)$ and $(t_5,~t_6)$ the magnetic gradient is $b(t)=+\eta_\text{b}$ while in $(t_2,~t_3)$ and $(t_4,~t_5)$ the gradient is switched to $b(t)= -\eta_\text{b}$. Due to the superposition of the spin ($S^\text{L,R}=\pm 1$), the CoM wave packets experience opposite splitting forces due to the NV. Thus in the left (red and light dash lines) and right (blue and solid lines) arms of interferometer CoMs possess opposite accelerations (donated by $a^{\text{L,R}}$), velocities and trajectories (denoted by $X^{\text{L,R}}$). The duration time for acceleration and deceleration are the same as $\tau_\text{a}$. We also assume a free-flight process between $t_3$ and $t_4$ with duration $2\tau_\text{f}$ where the magnetic gradient is turned off. Therefore, the entire time for the diamond flying inside the Stern-Gerlach apparatus is $4\tau_\text{a}+2\tau_\text{f}$.}
   }
		\label{figure2}
	\end{figure*}

	We do not consider the motion of the diamond in other dimensions, such as in drop tower experiments where the diamond needs to move in the direction of Earth's acceleration or in magnetic levitation setups where it moves along a direction orthogonal to $\mathbf{e}_x$. In these cases, the form of a magnetic field becomes very complicated for a closed loop, and it needs a careful design, especially when one takes the diamagnetic term into account \cite{Run1,Zhou:2022jug}. For simplicity, we consider the movements of CoMs only along $\mathbf{e}_x$, and they are subjected only to the spin-magnetism interaction (second term in Eq.~(\ref{CoM})). Therefore, to get a closed loop, the magnetic gradient $b(t)$ can be expressed in a very simple form as~\cite{Toros:2020dbf}, 
	\begin{equation}
		b(t)=\\
		\left\{
		\begin{array}{lr}
			+\eta_\text{b},~~&t=[t_1,t_2]\\\\
			-\eta_\text{b},&t=[t_2,t_3]\\\\
			0,&t=[t_3,t_4]\\\\
			-\eta_\text{b},&t=[t_4,t_5]\\\\
			+\eta_\text{b},~~&t=[t_5,t_6]\\
		\end{array}
		\right.\label{Protocol Acceleration}
	\end{equation}
	where in different time intervals, the gradients take constant values $\pm\eta_\text{b}$. The closed loop is guaranteed as long as the acceleration and deceleration durations (denoted as $\tau_\text{a}$) are the same. In the interval $[t_3,~t_4]$ there is a ``free-flight'' process with duration $\tau_\text{f}$, where the gradient is turned off, as illustrated in Fig. \ref{figure2} (a).

 In this paper, we concentrate on the contrast loss of phonon modes, while the mismatch issue in the CoM position and momentum of the two paths of the superposition is beyond the scope of this work. The maximum accelerations of the two paths are thus a series of constants given by,
	\begin{equation}
		\begin{aligned}
			a^\text{L,R}(t)=\frac{\mu S^{\text{L,R}}b(t)}{M}\,,
		\end{aligned}\label{acceleration}
	\end{equation}
	we assume that at the initial moment, the spin inside the diamond has been prepared in a superposition, and the CoM of the diamond is at rest. When the loop starts, wave packets on either sides of the interferometer behave distinctly according to $S^\text{R}$ and $S^\text{L}$, with opposite accelerations as illustrated in Fig. \ref{figure2} (b). In this model, we neglect the acceleration due to the induced diamagnetic potential for the time being. In fact, including the effect will not modify the current scenario a lot, as what matters is the relative force difference between the left and the right trajectories of the interferometer.

	In this simplest case, the CoM trajectories are symmetric given by $X_\text{c}^\text{R}=- X_\text{c}^\text{L}$, as depicted in Fig. \ref{figure2} (d) they can be expressed by the following expressions,
	\begin{equation}
		X_c^\text{L,R}(t)=\\
		\left\{
		\begin{array}{lr}
			\frac{a^\text{L,R}\left(t+2\tau_\text{a}+\tau_\text{f} \right)^2}{2},~~&t=[t_1,t_2]\\\\
			\frac{a^\text{L,R}\left(t^2-2\tau_\text{a}^2+2\tau_\text{f}t+\tau_\text{f}^2\right)}{2},&t=[t_2,t_3]\\\\
			a^\text{L,R}\tau_\text{a}^2,&t=[t_3,t_4]\\\\
			\frac{a^\text{L,R}\left(t^2-2\tau_\text{a}^2-2\tau_\text{f}t+\tau_\text{f}^2\right)}{2},&t=[t_4,t_5]\\\\
			\frac{a^\text{L,R}\left(-t+2\tau_\text{a}+\tau_\text{f} \right)^2}{2},~~&t=[t_5,t_6],
		\end{array}
		\right.\label{trajectories}
	\end{equation} 
    Thus, the relative separation distance $\Delta X(t)$ between the two trajectories is also a quadratic function of time. During the time interval $[t_3, t_4]$, this distance between the two trajectories reaches the maximum value,
	\begin{equation}
		\Delta X_\text{m}=2\left|a^\text{L,R}\right|\tau^2_\text{a},\\
	\end{equation} 
	where we define the positive constant maximum acceleration $\left|a^\text{L,R}\right|=\mu \eta_\text{b}/M$, and total duration time $\Delta t = t_6-t_1$.

	%%%%%%%%%%%%%%%%%%%%%%%%%%%%%%%%%%%%%%%%%%%%%%%%%%%%%%%%%%%%	
	%%%%%%%%%%%%%%%%%%%%%%%%%%%%%%%%%%%%%%%%%%%%%%%%%%%%%%%%%%%%
 \subsection{Spin-Phonon Interaction }
	So far, we can observe from the Hamiltonian (\ref{CoM}) that the diamond experiences two types of forces, i.e., the separation force acting on the NV particle located at $X_\text{c}$, and the diamagnetic force exerted by the entire diamond, which also effective located at CoM. Regarding the former, the force acts only on the central particle, the spin, which can be read out by doing partial derivative to the second term in Eq.~(\ref{CoM}). The splitting force has the form,
	\begin{equation}
		\begin{aligned}
			F^\text{L,R}_\text{s}(t)=\mu S^\text{L,R}b(t),
		\end{aligned}\label{FSpin}
	\end{equation}
	where the index ``s'' represents the spin. From here, one can immediately write down the spin-magnetic field-phonon interaction as $F_\text{s} x_\text{s} $ (recall that $x_\text{s}$ represent small vibration of the NV away from its equilibrium position). For clarity, in the following discussion relating to phonon, we will call this coupling as spin-phonon (Sp-Ph) interaction.
	\subsection{Diamagnetism-Phonon Interaction}
    
    For the diamagnetic force acting on the diamond, we consider that each particle has different magnetization opposite to the direction of the external magnetic field depending on its equilibrium position $X_i$. Therefore, each atom generates different magnitude of diamagnetic repulsion $F_\text{dia}^{\text{L,R}}(X_i):=F_{i,\text{dia}}^{\text{L,R}}$. From a macroscopic perspective, the overall diamagnetic repulsion experienced by the diamond is contributed by these atoms, thus we should obtain the following relationship (details of this part can been found in Appendix. \ref{NN}), 
	\begin{equation}
		\begin{aligned}
            F_\text{c}^\text{L,R}=\frac{\chi_\rho M}{\mu_0}\left(B_0b+b^2 X_\text{c}^\text{L,R}\right) =  \sum_i F_{i,\text{dia}}^\text{L,R}.
		\end{aligned}\label{Fdia}
	\end{equation}
	where the second part of (\ref{Fdia}) comes directly from the third term of Eq. (\ref{CoM}), representing the overall repulsion effectively exerted on the CoM. Therefore, the coupling between magnetic field (diamagnetism) and $i^\text{th}$ particle (phonon) in the chain can now be written as $F_{i,\text{dia}}^\text{L,R}x_i$. In the following discussion relating to phonon, we will call this coupling as diamagnetism-phonon (Dia-Ph) interaction.

	Recalling that we used a linear magnetic field, this simple magnetic field is widely applied in interferometers~\cite{Henkel_2019,Marshman:2021wyk}. Since the magnetic field gradient can be modelled as a step function, the separation force Eq.~(\ref{FSpin}) acting on the spin is independent of its equilibrium position $X_\text{s}$. Therefore, there is no coupling between the trajectories of wave packets and the phonon degrees of freedom. 
	
	In contrast to the splitting force acting on the NV centre, the situation for the diamagnetic force (\ref{Fdia}) acting on each particle in the atomic chain is more complicated. The tiny vibrations of these particles (phonons) are coupled to their equilibrium positions (trajectories) through the magnetic field (considering $F_{i,\text{dia}}^\text{L,R}\times x_i$ for each particle). Additionally, as shown in Fig. \ref{figure1} (a), the forces acting on the NV centre in different wave packets are of the same magnitude but in opposite directions. Therefore, when the loop is completed, the contrast of phonons modes always differs by a positive or negative sign in both normal coordinates and momentum. For the diamagnetic force in Fig. \ref{figure1} (b), the corresponding atoms in the atomic chain of the two wave packets experience the same direction of diamagnetic force but with different magnitudes (depending on the scale of spatial superposition). It is not difficult to imagine that the phonon modes will exhibit contrast reduction under such differential influences.

	\begin{figure*}[t!]
		\begin{subfigure}[]{0.38\textwidth}
			\includegraphics[width=\textwidth]{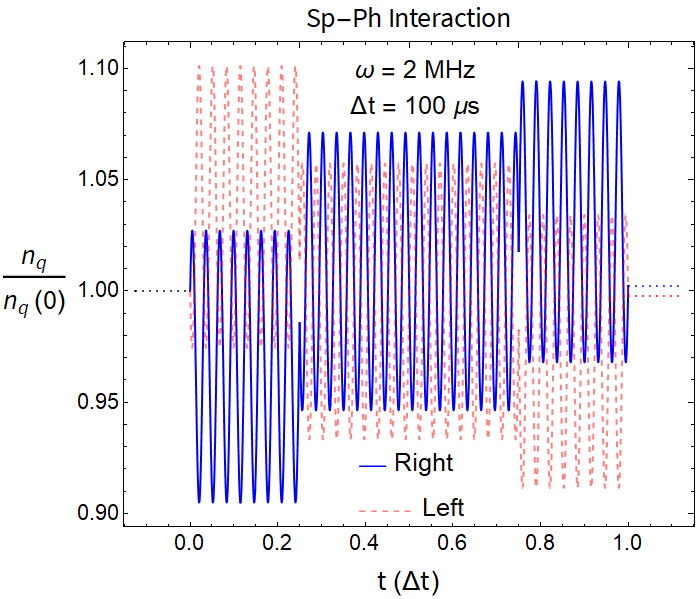}
			\subcaption*{(a)}
		\end{subfigure}
  \quad \quad \quad
		\begin{subfigure}[]{0.38\textwidth}
			\includegraphics[width=\textwidth]{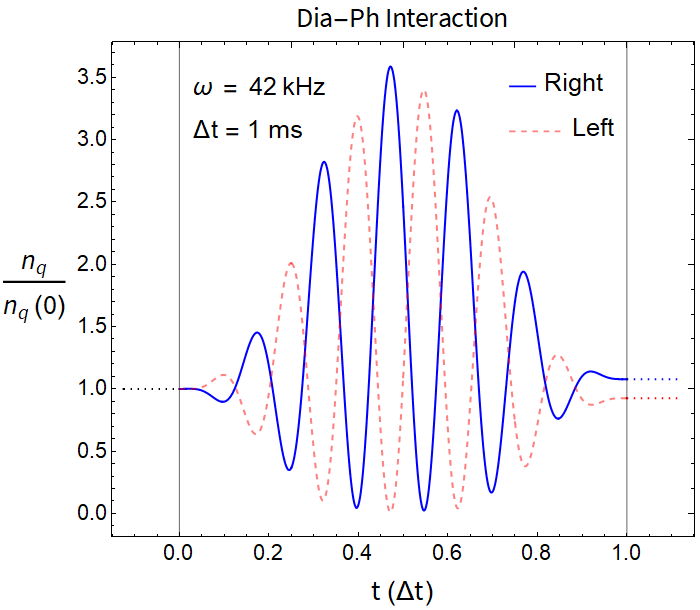}
			\subcaption*{(b)}
		\end{subfigure}
		\caption{
			\small {
				Illustrations of occupation numbers of the phonons for spin-phonon interaction (panel (a)), and for the the diamagnetism-phonon interaction (shown in panel (b)). The blue and light-red dash curves represent the wave packets of the diamond on the right and left sides, respectively. The black dot line represents the number of phonons in the diamond before entering the SG device (scaled to one), and the blue and red dot lines represent the number of phonons on both sides after the closure of the trajectory of the diamond's CoMs. Before the diamond enters the Stern-Gerlach interferometer, its wave packet is not split, so the initial number of phonons in the $q$ mode is determined only by temperature Eq. (\ref{nq0}). After it enters the Stern-Gerlach interferometer, the number of phonons on both sides of the interferometer arms changes over time and a difference in quantity forms. The phonon frequencies shown here (2 MHz for (a) and 42 kHz for (b)) are very low, in order to achieve better visibility. For the diamond we are interested in (specifically $10^{-14}$ kg), its natural frequency reaches $10^{10}$ Hz. Similarly, at those frequencies, the number of phonons on both sides of the interferometer arms also forms a difference. However, since the frequency of the magnetic field variation (especially the $\sim 1$s required for the QGEM experiment) is much lower than the phonon frequencies, the change in the number of phonons in those frequency ranges is not drastic.}} 
		\label{figure3}
	\end{figure*}
	
	%%%%%%%%%%%%%%%%%%%%%%%%%%%%%%%%%%%%%%%%%%%%%%%%%%%%%%%%%%%%%%%%%%%%%%
	
	\section{Phonon fluctuations}\label{secPhonon}
	
	In this section, we study phonon excited by the two forces~(\ref{FSpin}, \ref{Fdia}), acting on the lattice chain on the contrast of phonon wave packets. Here, we generally represent the aforementioned forces as  $F_i(X_i^\text{L,R})$ (we will neglect $X_i^\text{L,R}$ for simplicity sometimes and use just $F_i^\text{L,R}$). The total Hamiltonian regarding every atom and their interactions with external fields can be written as, 
	\begin{equation}
		\begin{aligned}
			H_\text{tot}&= H_\text{free}-\sum_i F_i^\text{L,R} x_i\\
			&= H_\text{free}-\sum_{i,q} \frac{Q_q^i F_i^\text{L,R}}{\sqrt{M}} u_q,
		\end{aligned}\label{PhononTotal}
	\end{equation}
	where $H_\text{free}$ is the free phonon Hamiltonian mentioned in Eq.~(\ref{Free}). 
	Eq.~(\ref{PhononTotal}) represents the energy change of the diamond interacting with the magnetic field after entering the SG apparatus. 
	
	When the diamond enters the SG apparatus, the interaction between its internal particles and the external field changes the number of phonons. With the Hamiltonian Eq.~(\ref{PhononTotal}), the kinematic equations for the canonical coordinates of each vibrational mode $q$ can be derived as,
	\begin{equation}
		\ddot{u}_q+\omega_q^2 u_q:=f^\text{L,R}_q=\sum_i\frac{Q_q^i F_i^\text{L,R}}{\sqrt{M}}\,, \label{13}
	\end{equation}
	where $f_q$ encompasses the forces acting on all atoms, it can be regarded as the force acting on the lattice chain. The solution of Eq.~(\ref{13}) is,
	\begin{equation}
		\begin{aligned}
			u_q(t)= & u_q(0) \cos (\omega_q t)+\dot{u}_q(0) \frac{\sin \left(\omega_q t\right)}{\omega_q} \\
			& +\int_0^{ t} \frac{\sin \omega_q (t-t^\prime)}{\omega_q} f^\text{L,R}_q(t^\prime) \mathrm{~d} t^\prime\,.
		\end{aligned}\label{3-2}
	\end{equation} 
	Here, the initial conditions, the phonon amplitude $u_q(t=0)$, and its conjugate momentum $\dot{u}_q(t=0)$ are solely dependent on the initial temperature of the diamond Eq.~(\ref{12}). The initial phonon population in mode $q$ is (\ref{nq0}) for the thermal equilibrium mixed state considered here.
	From here, it can be observed that due to the difference in forces acting on the atomic chains on the two sides of the interference arm, a discrepancy exists in the population of phonons with mode $q$.
	\begin{equation}
		\begin{aligned}
			n_q(t)  = \frac{1}{2\hbar \omega_q}\left(\dot{u}^2_q + \omega_q^2 u_q^2 \right).
		\end{aligned}\label{PhononNumber}
	\end{equation}
	Recall that $F_i^\text{L,R}$ in (\ref{13}) is a generalized form introduced for convenience. When investigating spin-phonon coupling, it needs to be substituted with (\ref{FSpin}). In this case, the summation over atomic index $i$ disappears (we consider only a single NV-centre). When studying the influence of the diamagnetic force of the lattice chain in different trajectories on internal degrees of freedom, $F_i^\text{L,R}$ is replaced by (\ref{Fdia}), and the summation in (\ref{13}) should include all atoms. 
	
	In Fig. \ref{figure3}, we have separately illustrated the differences in the number of phonons induced by these two types of forces (\ref{FSpin},~\ref{Fdia}). Recall that the eigenvector $Q_i^q$ of phonons is a quantity related to the atom index $i^\text{th}$. Here, we choose it to be one as the upper limit. the number of phonons is constant at the initial moment. After the diamond enters the SG apparatus, the number of phonons in the two trajectories changes differently and stabilizes with a constant difference after the duration time $\Delta t$. Recall that we are interested in diamond masses ranging from $M\sim 10^{-18}$ to $10^{-14}$ kg with frequencies $\omega_q$ in the range of $\omega_q\sim 10^{10}- 10^{11}$ Hz. Therefore, the rate of change of the population of phonons of mode $q$ is extremely rapid. Here, for visual clarity, we plot the variation in the number of low-frequency phonons in two types of interactions. One should note that due to the size constraints of diamonds ($\omega_0= \pi c/L$), elastic waves in the material do not reach such low frequencies.
	
	Here, we introduce the difference in phonon amplitude and momentum on both sides of the interferometer arms, which arise from $f_q^\text{L,R}$.
	\begin{equation}
		\begin{aligned}
			& \Delta u_q=\int_0^{t} \Delta f_q(t^\prime) \frac{\sin \omega_q (t-t^\prime)}{\omega_q} \mathrm{~d} t^\prime\\
			& \Delta \dot{u}_q=\int_0^{t} \Delta f_	q(t^\prime) \cos \omega_q (t-t^\prime) \mathrm{d} t^\prime,
		\end{aligned}\label{difference}
	\end{equation}
	where we define $\Delta f_q= f_q^\text{R}-f_q^\text{L}$.
	From here, one can observe that due to the spatial superposition of CoM's,  the lattice chain in the two wave packets experiences different forces. In Fig. \ref{figure4}, we illustrate the differences in the phase space of phonon wave packets from different paths. For visual clarity, the early stage of phase evolution is faded out, and normal coordinate and its conjugate have been made dimensionless.  In \cite{Henkel1,Henkel2}, the authors have shown that the phase space trajectories of phonons exhibit differences because the spin in the lattice is moving toward different directions, as shown in Fig. \ref{figure4} (a). In the case of diamagnetism-phonon interaction, the differences in the force value exerted on the lattice still contribute to the reduction in contrast for phonon wave packets. In the next section, we will estimate the loss of contrast due to their couplings.

	%%%%%%%%%%%%%%%%%%%%%%%%%%%%%%%%%%%%%%%%%%%%%%%%%%%%%%%%%%%%%%%%%%%%%%%%%%%%
	
	\section{Loss of Contrast}\label{Loss of Contrast}
	The contrast in the internal degrees of freedom is defined as the inner product of the quantum states of phonons after evolving in the two interferometry arms; for a specific mode $q$, it becomes,
	\begin{equation}
		C_q=|\langle\psi^\text{L}_q(\Delta t)|\psi^\text{R}_q(\Delta t)\rangle|. \label{18}
	\end{equation}
	From Eq.~(\ref{18}), it can be seen that the contrast is essentially the inner product of normalized phonon wave functions. Thus, a perfect closure of position, momentum, and other relevant degrees of freedom, including phonon vibrations, would yield $C_q=1$, while an imperfect closure results in $C_q \leq 1$ indicating the loss in contrast.
	We employ the time-dependent perturbation theory in quantum mechanics for the time evolution of phonon quantum states. In Eq.~(\ref{PhononTotal}), the interaction between phonons and external field in the equation is considered as a small perturbation. Now, we ignore the summation over $q$ and focus on one mode; the Hamiltonian becomes, 
	\begin{equation}
		\begin{aligned}
			H_q &= \hbar \omega_q a_q^\dagger a_q - \sum_i F_i^\text{L,R} x_i\\
			&= \hbar \omega_q a_q^\dagger a_q - \sum_i F_i^\text{L,R} Q_q^i\sqrt{\frac{\hbar}{2M \omega_q}}(a_{-q}^\dagger+a_q).
		\end{aligned}\label{hamiltonianq}
	\end{equation}
	Here, recall that the force $F_i^{\text{L,R}}$ is time dependent and it is subjected to the equilibrium positions $X^\text{L,R}_i(t)$ of the atomic chain. Considering that the diamond is in a spatial superposition state, the forces acting on the atomic chains in the two trajectories are different. 
	In the interaction picture, denote the time-dependent perturbation part in Eq.~(\ref{hamiltonianq}) as $V_q(t)$, the quantum state of phonon obeys the Schrödinger equation. The overlap of phonon modes in the two arms becomes,
	\begin{equation}
		\begin{aligned}
			&	\langle\psi_q^\text{L}(\Delta t)|\psi_q^\text{R}(\Delta t)\rangle\\
			&=\langle\psi_q(0)|\exp\left\{\frac{\text{i}}{\hbar}\int_0^{\Delta t} \text{d}t^\prime \left[V^\text{L}_{q}(t^\prime)-V^\text{R}_{q}(t^\prime)\right]\right\}|\psi_q(0)\rangle\\
			&= \langle\psi_q(0)|U_q^{\text{L},\dagger}U_q^\text{R}|\psi_q(0)\rangle,
		\end{aligned}\label{hamiltonianq3}
	\end{equation}
	where $U_q$ is the time evolution operator, 
	\begin{equation}
		U_q^{\text{L},\dagger}U_q^\text{R}=\exp \frac{\text{i}}{ \hbar}\left[\Delta \dot{u}_q \hat{u}_q+\Delta u_q \hat{\dot{u}}_q\right], \label{displacement}
	\end{equation}
	when computing the second line of Eq.(\ref{hamiltonianq3}), $\Delta \dot{u}_q(t)$ and $\Delta u_q(t)$ emerge, which are precisely defined in Eq.(\ref{difference}). 
	%%%%%%%%%%%%%%%%%%%%
	\begin{figure*}[t!]
		\begin{subfigure}[]{0.38\textwidth}
			\includegraphics[width=\textwidth]{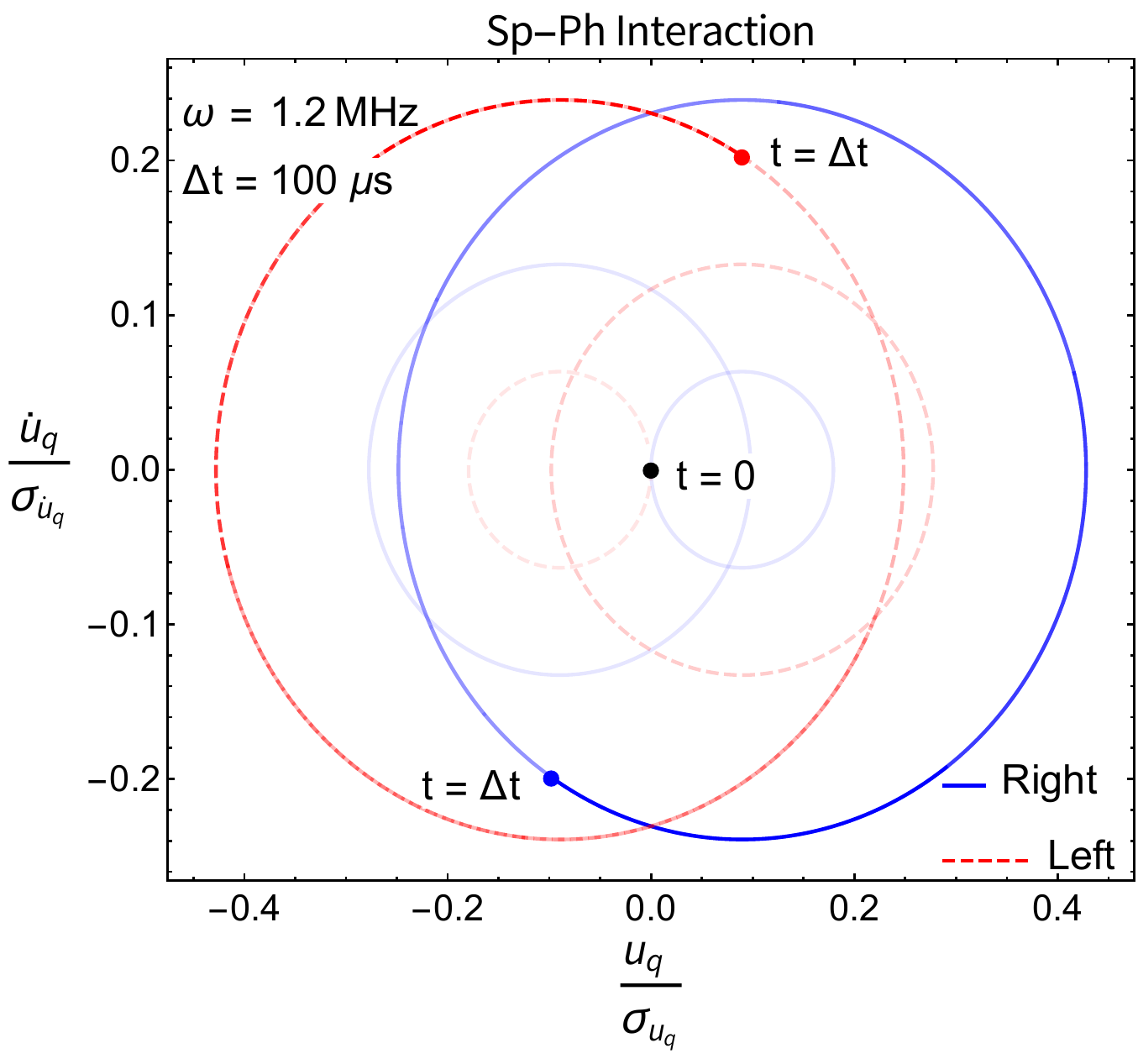}
			\subcaption*{(a)}
			\label{spin1}
		\end{subfigure}
  \quad \quad \quad
		\begin{subfigure}[]{0.38\textwidth}
			\includegraphics[width=\textwidth]{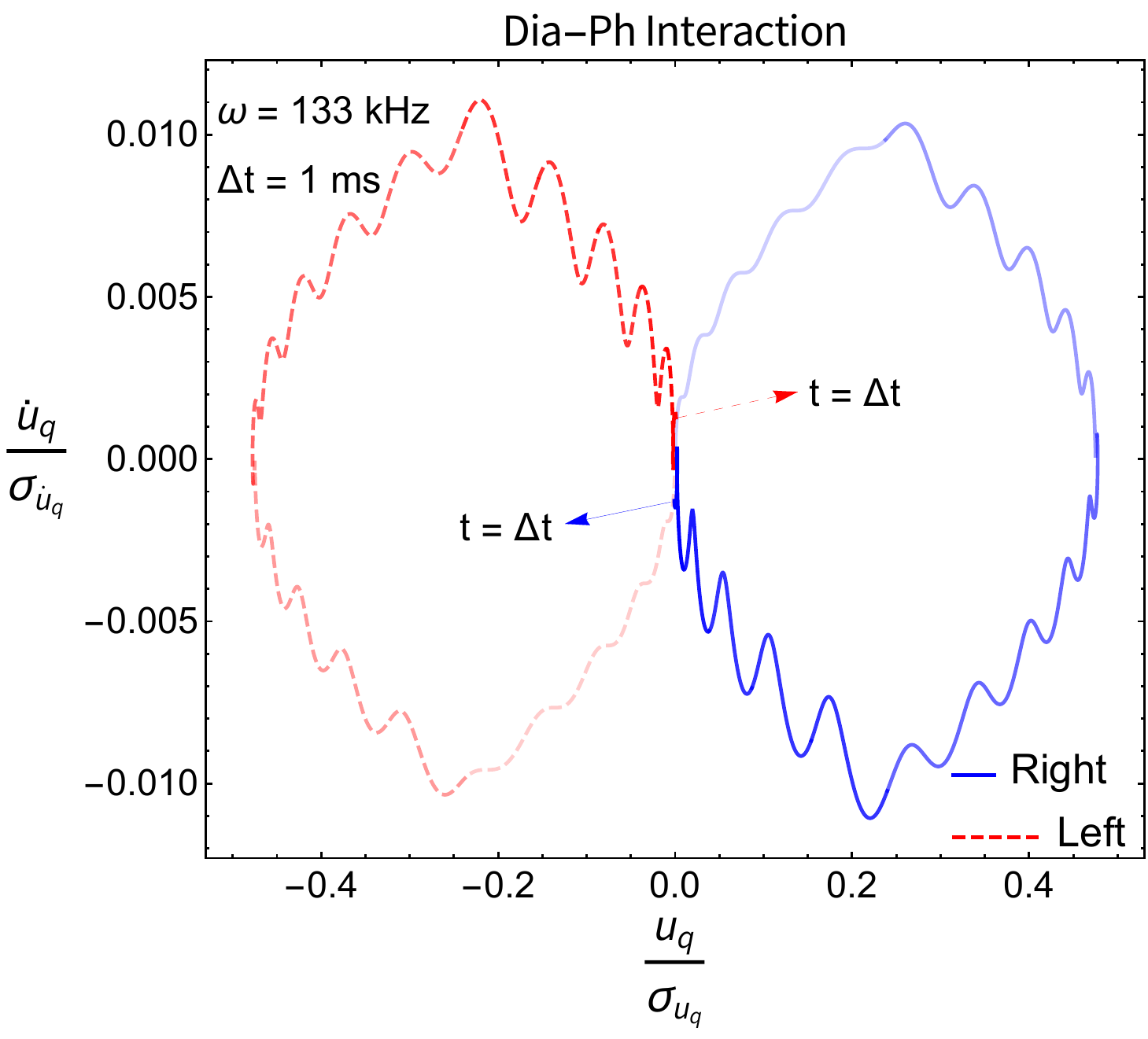}
			\subcaption*{(b)}
			\label{dia1}
		\end{subfigure}
		\caption{
			\small {
				Splitting in the phase space of a fixed phonon mode where the horizontal and vertical axis represent phonon amplitude and momentum, respectively. Panel (a) represents the difference induced by Sp-Ph interaction, and panel (b) shows the differential phase due to Dia-Ph interaction. The blue and light-red dash curves, respectively, represent the wave packets of the diamond on the right and left sides. The early stages of the phase space trajectories are faded for clarity.}} 
		\label{figure4}
	\end{figure*}
	%%%%%%%%%%%%%%%%%%%%%%%%%%%%%%%%	
	In Eq.~(\ref{displacement}), one can first consider the case of spin-phonon coupling (\ref{FSpin}). In this case, $V_q^\text{L}$ and $V_q^\text{R}$ differ exactly by a positive or negative sign, which is because the forces exerted on the phonon wave packets on both sides of the interferometer arms happen to be in opposite directions. In this scenario, the amplitudes and momentum of the phonons on both sides of the interferometer arms are displaced in opposite directions by $\frac{1}{2}\Delta u_q$ and $\frac{1}{2}\Delta \dot{u}_q$, respectively. The contrast of phonon can be comprehended as the probability of its amplitude and momentum staying in the initial state $\psi_q(0)$~\cite{Henkel1} as shown in Fig. \ref{figure4} (a), once the two wave packets are separated, the overlap would not be perfect.

	For the diamagnetism-phonon interaction, the decrease in contrast is also not difficult to imagine. In this case, the amplitudes and corresponding momentum of the phonons on both sides of the interferometer arms are displaced to varying degrees in the same direction. One can always utilize Eq.(\ref{hamiltonianq3}) to calculate the contrast of the phonon wave function. (details are given in the appendix) as,
	\begin{equation}
		\begin{aligned}
			& C=|\left\langle\psi^{\mathrm{L}}(\Delta t) \mid \psi^{\mathrm{R}}(\Delta t)\right\rangle| \\
			& =\prod_q \left| \text{Tr}\left[U_q^{\text{L},\dagger}U_q^\text{R} |\psi_q(0)\rangle\langle\psi_q(0)|\right]\right|\\
			& =\prod_q \left| \text{Tr}\left[U_q^{\text{L},\dagger}U_q^\text{R} \rho_q(0)\right]\right|.
		\end{aligned}\label{Contrast}
	\end{equation}
	Here, we have considered an initial thermal equilibrium ensemble of phonons composed of a state $\rho_q(0)$.
	
	Note that to evaluate the contrast Eq.~(\ref{Contrast}), it is equivalent to evaluate the mean value of $U_q^{\text{L},\dagger}U_q^\text{R}$
	which consists of both phonon’s “position” and its conjugate. Therefore, it is convenient to calculate the mean value by using the Wigner function, which is a quasi-distribution of phonon's position and momentum in phase space, where for a harmonic mode considered here, it has a form of double Gaussian Eq.~(\ref{Wigner}).
	
	Therefore, in the end, by multiplying the exponential operator with Eq.~(\ref{Wigner}) and integrating over $u_q$ and $\dot{u}_q$, the contrast becomes,
	\begin{equation}
		\begin{aligned}
		C_q=	& \text{Tr}\left[U_q^{\text{L},\dagger}U_q^\text{R} \rho_q(0)\right]\\
			&= \exp \left[-\frac{\operatorname{coth} \frac{1}{2} \frac{\hbar \omega_q}{k_B T}}{4 \hbar \omega_q}\left|\int_0^{\Delta t} \Delta f_q(t) \mathrm{e}^{\text{i} \omega_q t} \mathrm{~d} t\right|^2\right]\\	
			&= \exp \left[-\frac{\operatorname{coth} \frac{1}{2} \frac{\hbar \omega_q}{k_B T}}{4 \hbar \omega_q}\left|\Delta \tilde{f}_q(\omega_q)\right|^2\right].\\
		\end{aligned}\label{Contrastnew}
	\end{equation}
	The final result of the contrast is an exponential function and is temperature-dependent. Naturally, when $\Delta \tilde{f}_q(\omega_q)$ is zero, interference arms on both sides will exhibit no distinction, and Eq.~(\ref{Contrastnew}) will yield a result of one. However, if there is a difference in the forces acting on the lattice chain on both sides of the interference arms, then $C \neq 1$, and the loss of contrast~\footnote{From the expression of contrast, one can directly see why high temperature will destroy contrast. Suppose $T\rightarrow \infty$ for a fixed $\omega_q$ then the expression for the exponential part of 
 Eq.(\ref{Contrastnew})
 goes as, $-\frac{k_\text{B}T}{2\hbar^2\omega_q^2}|\Delta\tilde{f}_q|^2\rightarrow -\infty$, therefore $C\rightarrow 0$. The transfer function $\Delta\tilde{f}_q$ is always non-zero in our case.
 In the other limit, when $\omega_q\sim cq\rightarrow 0$, for a finite temperature, again means that the $C\rightarrow 0$. Lower values of $\omega_q $ can occur for systems with a lower speed of sound. Typically, these are more compressible objects for which the speed of sound is smaller than that of the diamond-like system. }is expressed as $\ln C$. The Fourier transform,
	\begin{equation}
		\begin{aligned}
			\Delta \tilde{f}_q:=\Delta \tilde{f}_q(\omega_q)=\left|\int_0^{\Delta t} \Delta f_q(t) \mathrm{e}^{\text{i} \omega_q t} \mathrm{~d} t\right|,
		\end{aligned}\label{TransformFunction}
	\end{equation}
	represents a power spectrum that is a function of $\omega_q$ and $\Delta t$, determining the value and properties of the contrast. In the study of noise in SG devices, macroscopic noise is transformed into the phase fluctuation through this transform. Thus it is also referred to as the transfer function \cite{Wu:2022rdv} (see also \cite{Toros:2020dbf}). The transfer function is closely related to the trajectory of the interference arms. For the simplest acceleration protocol (Fig. 1), the transfer function of the splitting force is very trivial, transforming two constant accelerations. The resulting $f_q(\omega_q)$ for this transformation only contains $1/\omega_q$. However, for forces closely related to the trajectory, their transfer function contains powers of $(1/\omega_q)^n$. In the following sections, we will compute the contrast.
	
	%%%%%%%%%%%%%%%%%%%%%%%%%%%%%%%%%%%%%%%%%%%%%%%%%%%

	\section{Numerical Results}\label{Results}
	
	In this section, we numerically explore contrast reduction due to phonon wave packets from the interactions given in Eq.~(\ref{CoM}). Since we do not want the magnetic field gradient, which controls the spatial superposition size, to be too large (i.e., $\eta_\text{b} \le 10^{6}$ T/m, see such gradients can be achieved in~\cite{machluf2013coherent,Mamin2012HighFD}), thus
    the maximum spatial separation distances $\Delta X_\text{m}\sim 10^{-6}-10^{-3}$m for different masses. For simplicity, here we consider phonons as longitudinal elastic waves,   
	$\omega_q=c q$~\cite{jishi2013feynman}, and treat them as continuous with a minimum frequency (fundamental frequency) $\omega_0={\pi c}/{L}$. Besides, we take the eigenvector $Q_q^i=1$ (recall that in solid state physics, these vectors are sinusoidal functions) as an upper limit. We also define the moment when the diamond wavepacket reaches its maximum separation distance $\Delta X_\text{m}$ as time $t=0$. Thus, $t_1 = -2\tau_\text{a} - \tau_\text{f}$, and $t_6 = -t_1 = 2\tau_\text{a} + \tau_\text{f}$.
	
	To compute Eq.~(\ref{Contrastnew}), we need to convert Eq.~(\ref{FSpin}) and Eq.~(\ref{Fdia}) into respective transfer functions. Here, we first provide their forms in the temporal domain as differences on both sides of the interferometer arms,
	\begin{equation}
		\begin{aligned}
			\Delta F_s(t) &=\mu(S^\text{R}-S^\text{L})\eta_\text{b}
			=2\mu\eta_\text{b},\\
			\Delta F_{\text{dia}}(t) &= \sum_i \left(F_{i,\text{dia}}^\text{R}- F_{i,\text{dia}}^\text{L}\right)\\
			&= \frac{\chi_\rho M}{\mu_0}\eta_b^2\left(X_c^\text{R}-X_c^\text{L}\right)
		\end{aligned}\label{differential force}
	\end{equation}
	The differential force for spin-phonon interaction is easily understood in (\ref{differential force}). These are the constant differences between the left and right interferometer arms at different time stages. Recall that $\eta_\text{b}$ is the maximum value of the magnetic field gradient. 
	
Likewise, for the diamagnetism-phonon coupling, due to the diamond wave packet being in a spatial superposition, there is also a difference of the diamagnetic repulsion generated by the atoms located on the two sides of the interferometer arms (for example, the $i^\text{th}$ atom with positions $X_i^\text{L,R}$). 

	Thereafter, we can obtain their corresponding transfer functions,
	\begin{equation}
		\begin{aligned}
			&|\Delta \tilde{f}_\text{s} (\omega_q)|^2=M \left(\frac{2 \Delta X_\text{m}}{\tau_\text{a}^2 \omega_q}\right)^2\times \Gamma^2(\omega_q)\\
			&|\Delta \tilde{f}_\text{dia} (\omega_q)|^2=M^5\left(\frac{\chi_\rho \Delta X_\text{m}^3}{2\mu_0 \mu^2 \tau_\text{a}^6 \omega_q^3}\right)^2\times \Gamma^2(\omega_q)\,.
		\end{aligned}\label{delta transform}
	\end{equation}

	In Sec. \ref{Interactions in SGI}, the coupling of the magnetic field with the spin generates the spatial superposition state of the diamond wave packet. Here, we parameterize the trajectory, acceleration, magnetic field gradient, and other parameters of the wave packet in terms of the acceleration duration $\tau_\text{a}$, mass $M$, and maximum separation distance $\Delta X_\text{m}$. The $\Gamma(\omega_q)$ here arises from the Fourier transform. For the two types of interactions, their forms are the same,
	\begin{equation}
		\begin{aligned}
			\Gamma(\omega_q)= \sin[(\tau_\text{f})\omega_q]&-2\sin[(\tau_\text{a}+\tau_\text{f})\omega_q]\\
			&+\sin[(2\tau_\text{a}+\tau_\text{f})\omega_q].
		\end{aligned}\label{transform}
	\end{equation}
	Substituting the corresponding transfer function in Eq.(\ref{delta transform}) into equation Eq.(\ref{Contrastnew}), we can obtain the expression for the contrast reduction brought about by these interactions,
 \begin{equation}
		\begin{aligned}
			-\text{ln}&~C_\text{spin}=\\
			&\sum_q\coth\left(\frac{1}{2} \frac{\hbar \omega_q}{k_\text{B} T}\right)\frac{M \Delta X^2_\text{m}}{\tau_\text{a}^4\omega_q^3\hbar} \Gamma^2(\omega_q),\\
		\end{aligned}\label{lnspin}
	\end{equation}	
 \begin{equation}
		\begin{aligned}
			-\text{ln}&~C_\text{dia}=\\
			&\sum_q\coth\left(\frac{1}{2} \frac{\hbar \omega_q}{k_\text{B} T}\right)\frac{\chi_\rho^2}{16 \mu_0^2 \mu^4\hbar}\frac{M^5 \Delta X^6_\text{m}}{\tau_\text{a}^{12}\omega_q^7} \Gamma^2(\omega_q).\\
		\end{aligned}\label{lndia}
	\end{equation}
\begin{figure}[h]
	\includegraphics[width=0.509\textwidth]{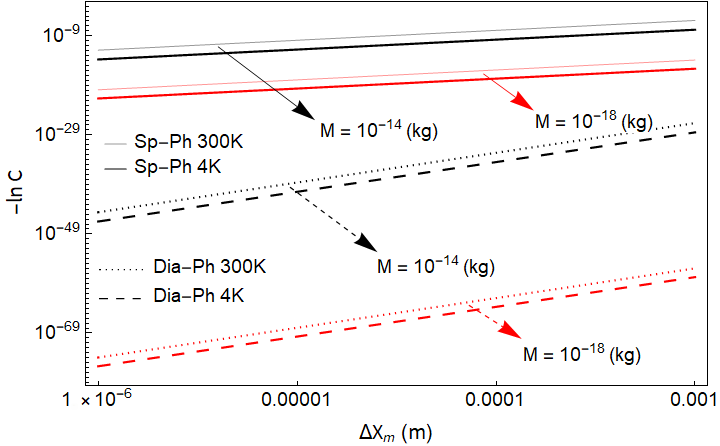}
	\caption{The figure shows the phonon-induced contrast loss caused by spin-phonon (solid lines) and diamagnetism-phonon (dash and dot lines) interactions in the expected scenario of the QGEM experiment (with a duration time of approximately one second). The horizontal axis $\Delta X_\text{m}$ is the maximum separation distance between diamond wave packets on either sides of the interferometer. From Eqs. (\ref{lnspin},~\ref{lndia}), the contrast loss caused by phonon modes will be zero when $\Delta X_\text{m}= 0$ indicating no spatial separation. In the figure, the contrast loss from heavier mass $M=10^{-14}$kg (four black lines) is correspondingly greater than that for the smaller mass $M=10^{-18}$kg (four red lines). The figure also depicts scenarios at different temperatures, showing that the phonon-induced contrast loss at 300K is greater than that at 4K.}
	\label{new}
\end{figure}

    Note that since we start from a one-dimensional atomic chain model, the mass $M$ in (\ref{lnspin}) and (\ref{lndia}) should, in principle, be understood as the mass of a single atomic chain within the diamond. However, in the subsequent calculations, we will interpret this $M$ as the total mass of an isotropic homogeneous cubic diamond with side length $L=(M/\rho)^{-\frac{1}{3}}$, where the main contribution to contrast loss comes from the fundamental tone $\omega_0=\pi c/L$.
    \footnote{Phonon modes depend on the shape of three-dimensional objects. Studies on acoustic phonons in circular objects can be found in \cite{PhysRevB.101.125404}. The situation becomes considerably more complex for arbitrary-shaped three-dimensional objects, and deserves separate discussion. }.

	In Fig. \ref{new}, we show the contrast reduction resulting from spin-phonon (solid lines) and diamagnetism-phonon (dash and dot lines) interactions in a setup with $\Delta t=1$s. The horizontal axis represents the maximum separation distance of the diamond's CoMs in the spatial superposition state. The black ($10^{-14}$ kg) and red ($10^{-18}$ kg) lines compare diamonds of different masses. We also compared the contrast loss at different temperatures,  \footnote{
		Since the main contribution of contrast loss comes from the fundamental frequency $\omega_0\sim c/L$. One can consider two limits here, one is $\hbar \omega_0\ll k_\text{B}T$ and the other is $\hbar \omega_0\gg k_\text{B}T$. For the former limit, the square of the transfer function will yield the maximum value of $|\Gamma(\omega_0)|^2\sim 1$, and Eqs.~(\ref{lnspin}) and (\ref{lndia}) become, respectively, 
		\begin{align*}
			-\text{ln}~C_\text{spin}\approx \frac{2k_\text{B}T}{\hbar^2}\frac{M \Delta X^2_\text{m}}{\tau_\text{a}^4\omega_0^4}\sim \frac{2k_\text{B} T }{ \hbar^2}\rho\left(\frac{ \Delta X_\text{m} }{c^2 \tau_a^2}\right)^2L^{7},
		\end{align*}
		\begin{align*}
			-\text{ln}~C_\text{dia}\approx \frac{k_\text{B}T\chi_\rho^2}{8\hbar^2 \mu_0^2 \mu^4}\frac{M^5 \Delta X_\text{m}^6}{\tau_\text{a}^{12}\omega_0^8}
			\sim \frac{k_\text{B}T\chi_\rho^2 \rho^5}{8 \mu_0^2\mu^4  \hbar^2}\left(\frac{\Delta X_\text{m}^3  }{c^4\tau_\text{a}^{6}}\right)^2L^{23}.
		\end{align*}
		In the other limit, when $\hbar \omega_0\gg k_\text{B}T$ the hyperbolic function goes $\coth\frac{1}{2}\frac{\hbar \omega_0}{k_\text{B}T}\rightarrow 1.$ The contrast reductions from the two interactions become as,
		\begin{align*}
			-\text{ln}~C_\text{spin}\approx \frac{M \Delta X^2_\text{m}}{\hbar\tau_\text{a}^4 \omega_0^3}\sim \frac{\rho}{\hbar c^3}\left(\frac{ \Delta X_\text{m} L^3}{\tau_\text{a}^2}\right)^2,
		\end{align*}
		\begin{align*}
			-\text{ln}~C_\text{dia}\approx \frac{\chi_\rho^2}{16\mu_0^2 \mu^4 \hbar}\frac{M^5\Delta X^6_\text{m}}{\tau_\text{a}^{12}\omega_0^7} \sim \frac{\chi_\rho^2}{16\mu_0^2 \mu^4 \hbar}\frac{\rho^5}{c^7}\left(\frac{\Delta X^3_\text{m}L^{11}}{\tau_\text{a}^{6}}\right)^2.
		\end{align*}
		In the above two limits, we have used the relation: $M=\rho L^3$, where $\rho$ is the density of test masses and $L$ is the length. In either of the limits, one can always conclude that more rigid material (which means faster sound speed) and smaller size of the test masses will produce less reduction in contrast. The diamond-like crystal will be an example of an ideal crystal for the QGEM experiment.} where the results at low temperature (4K) are presented by dark-solid lines and dash lines while at high temperature (300K) results are shown in light-solid and dot lines.
		The results indicate that the contrast loss induced by the spin-phonon coupling is dominant, while the effect of diamagnetism-phonon interaction is significantly smaller.  
		The source of these differences can be easily understood by comparisons in Eqs. (\ref{lnspin},~\ref{lndia}). The contrast reduction caused by the former interaction is suppressed by $\omega_q^3$, whereas the contrast loss caused by diamagnetism-phonon interaction is suppressed by $\omega_q^7$.

	\begin{figure*}[t!]
	\begin{subfigure}[]{0.45\textwidth}
		\includegraphics[width=\textwidth]{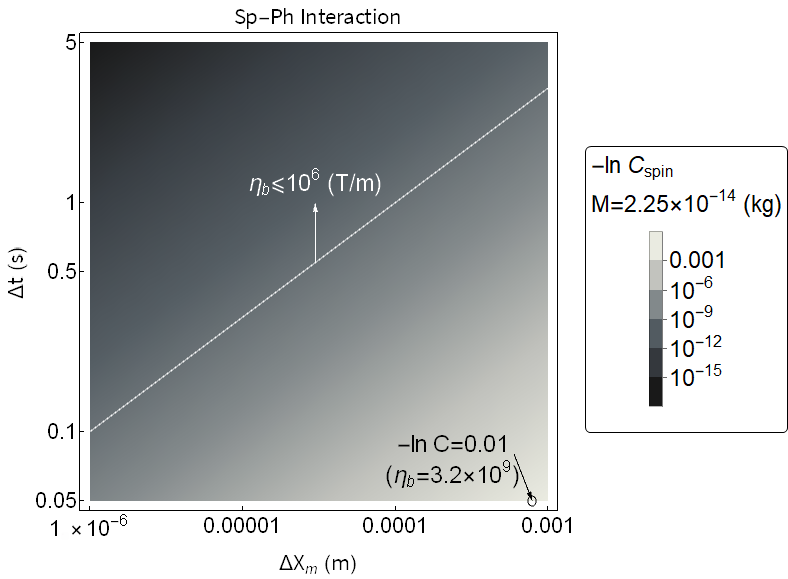}
		\subcaption*{(a)}
	\end{subfigure}
	\begin{subfigure}[]{0.45\textwidth}
		\includegraphics[width=\textwidth]{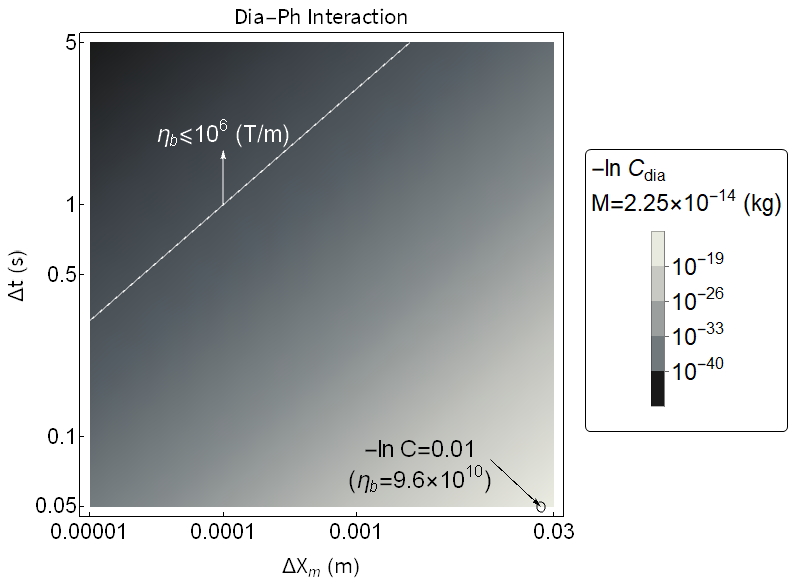}
		\subcaption*{(b)}
	\end{subfigure}
	\begin{subfigure}[]{0.45\textwidth}
		\includegraphics[width=\textwidth]{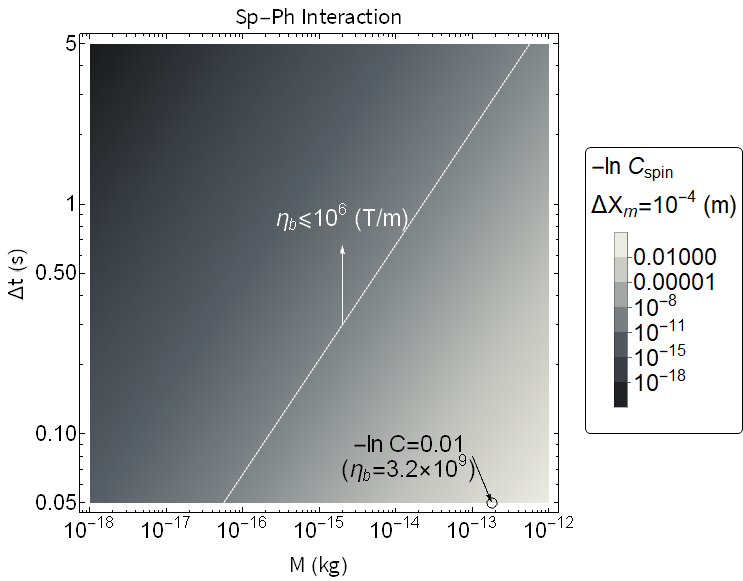}
		\subcaption*{(c)}
	\end{subfigure}
	\begin{subfigure}[]{0.45\textwidth}
		\includegraphics[width=\textwidth]{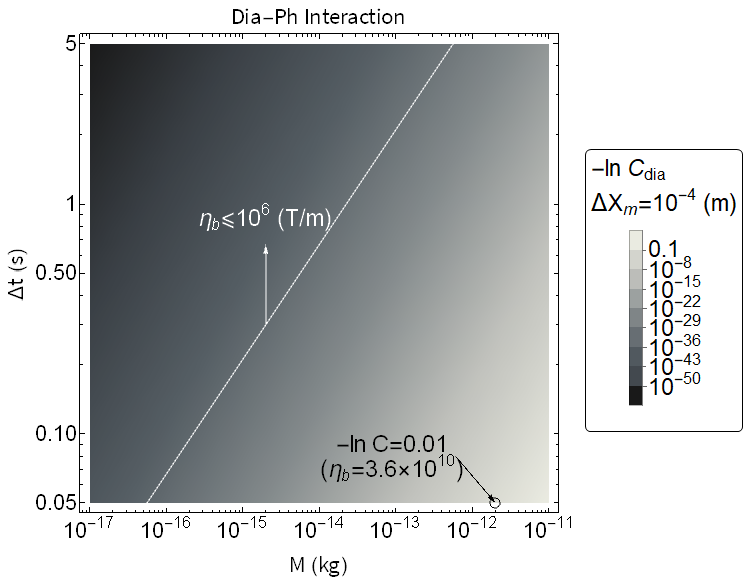}
		\subcaption*{(d)}
	\end{subfigure}
	\caption{
		\small {Panels (a), (b), (c), and (d) illustrate the contrast reduction resulting from spin-phonon coupling and diamagnetism-phonon interaction under different close-loop protocols, where the colour shading from dark to light represents progressively intensified contrast reduction. The first column corresponds to spin-phonon coupling, while the second corresponds to diamagnetism-phonon interaction. In the first row, (a) and (b) depict scenarios with a diamond mass of $M\sim 2.25\times 10^{-14}$ kg, showing different maximum separations $\Delta X_\text{m}$ (horizontal axis) and duration times $\Delta t$ (vertical axis). In the second row, (c) and (d) assume a desired separation distance of $\Delta X_\text{m}=10^{-4}$m for the diamond's CoMs. We explore the contrast loss caused by phonon wave packets of different masses under varying duration times.   
			In the four subfigures depicted, we have marked the cases where the contrast loss reaches $1\%$. In subfigures (a) and (b), for a diamond with mass $M=2.25\times10^{-14}$kg, significant contrast loss requires achieving maximum separation distances of approximately $\sim 10^{-3}$m and $\sim 10^{-2}$m for the CoM wave packet within 0.05 s, respectively. In subfigures (c) and (d), to achieve $\Delta X_\text{m}=10^{-4}$m within 0.05 s, the diamond mass needs to reach approximately $\sim 10^{-13}$kg and $10^{-12}$kg, respectively, to exhibit significant contrast loss. In these four cases marked, the magnetic gradients required are all large, becoming approximately $1000$ to $10000$ times greater than the current experimental capability.}}
	\label{figureX}
\end{figure*}

    Moreover, room temperature does not drastically alter the outcomes for either of them. Additionally, it can be observed that larger diamond masses lead to more phonon contrast reduction, which is quite understandable. Larger mass implies a lower range of phonon frequencies. From Eqs.~(\ref{Contrastnew}) and (\ref{delta transform}), it can be seen that these $\omega_q$ terms appear in the denominators.
	
	From Eq. (\ref{Contrastnew}), we can see that the acceleration protocol profoundly impacts the contrast of the phonon wave packets. In Fig. \ref{figureX}, we use a density plot to detail the contrast reduction resulting from different acceleration protocols. In the four subplots, we highlight the black curves, where the magnetic field gradient is precisely $10^6$ T/m, and in the region above it, the duration time is longer, thus requiring a smaller magnetic field gradient. In Fig. \ref{figureX} (a) and (b), we study the contrast of a diamond with a mass of $M=2.25\times 10^{-14}$ kg under different maximum separation distances $\Delta X_\text{m}$ and duration times. In (c) and (d), we set the CoM's maximum separation distance to $10^{-4}$ m which represent a sufficiently large separation distance of wave packets in QGEM and study the contrast loss for different diamond masses and duration times. 
 
 	From (a) and (b), one can see that for $10^{-14}$kg diamond, when the separation protocol requires the loop to be completed within a short time (for example, 0.05 s) while reaching large separation distance $\Delta X_\text{m}$, then the loss of contrast from both interactions could go to $1\%$. Likewise, in (c) and (d), for large spatial separation as expected by the QGEM (for example $10^{-4}$m), protocols with heavier diamond and shorter duration time $\Delta t$ would also lead to severe contrast reduction in phonon degrees of freedom. Overall, significant contrast loss will happen when wave packets of a large mass diamond are separated over a large spatial distance within a short period. However, within the current experimental constraints where the SG magnetic field is constrained to be less than $10^{6} ~ {\rm T/m}$~\cite{machluf2013coherent,Mamin2012HighFD}, the contrast loss induced by both couplings is negligible for QGEM.

	\section{Discussion}
	
	In this paper, we have analyzed the contrast loss due to phonon in creating spatial superposition of a nanoscaled diamond embedded with a single NV centre based on a Stern-Gaelach interferometer. 
	During the process of creating a spatial superposition state using the coupling between the spin and the linear magnetic field, the spin particle in the lattice chain generates a pair of forces (\ref{FSpin}) of opposite directions and equal magnitudes due to their own spin superposition state. Similar to the Humpty-Dumpty problem at the centre of mass of the diamond, the spin particles on both sides of the interferometer arm have different vibrational states in the lattice, resulting in differences in the quantum states of the phonons. This also leads to imperfections in the overlap of phonon wave packets.
	
	The other interaction arises from the diamagnetic properties of diamond. Each atom in the lattice chain produces a diamagnetic force (\ref{Fdia}) when subjected to an external magnetic field. This diamagnetic force is of the same direction but has a different magnitude on both sides of the interferometer arm. Similarly, due to this difference, the quantum states of phonons on both sides are also different, resulting in contrast reduction.

	In this paper, we use a linear magnetic field (\ref{BField}), which is widely used in atomic interferometers. Furthermore, this linearity greatly simplifies the calculation of the motion state of spin particles. That is, the splitting force is independent of trajectories and the equilibrium position. 
	
	The situation is more complex regarding the interaction between diamagnetic force and phonons. Since the diamagnetic term in the Hamiltonian includes $B(X_i)^2$, the force on the lattice is now coupled with its trajectory. This external field mediated coupling between phonon and CoM's trajectory is common in the study of optical cooling of the motion state of diamond centre. In this work, we typically studied the latter situation and made comparison of contrast reduction between these two interactions. 
	
	Specifically, we assumed that the phonons were initially at a specific temperature $T$, and we utilized the Wigner representation to express and compute the evolution of the quantum state of the phonons.

	Our results indicate that among these two sources contributing to the contrast loss in the phonon degrees of freedom, the effect of the spin-phonon coupling is dominant. The contrast loss induced by the diamagnetic term is smaller than the former, as shown in Fig. \ref{new}. This result can be directly understood from (\ref {lnspin}) and (\ref{lndia}). For the spin-phonon coupling, which does not depend on the trajectory, the contrast reduction is suppressed by the phonon frequency to the power of three. Meanwhile, the contrast loss due to diamagnetism-phonon interaction is suppressed by the phonon frequency to the power of seven. 
	
	The latter case directly arises from its coupling with the trajectory, as the diamagnetic force is a function of time, and its variation is adiabatic. In the seminal work \cite{Henkel1,Henkel2}, this is also referred to as the degree of adiabaticity.

	However, overall, the contrast reduction induced by both interactions is minimal. This is good news for matter-wave interferometers. There are two main reasons for this result. One is that we limit the gradient of the magnetic field. The most important reason, however, is that we are studying diamonds of very small size ($10^{-20}\sim 10^{-14}$ kg) compared to the speed of sound $c$ multiplied by the duration time (on the order of microseconds and 1 second for QGEM). 
 	Since the frequencies of the phonons are very high (tens of gigahertz to several terahertz), their contribution to contrast loss is tiny. However, if we prefer a different material where the speed of sound is much smaller than that of the diamond, we would see the loss of contrast occur. However, it also depends on how we create the superposition; hence, it depends on the details of the transfer function.

	Acknowledgements: SB thanks EPSRC grants EP/R029075/1, EP/X009467/1, and  ST/W006227/1. Q.X and R.Z are supported by the China Scholarship Council (CSC).

	\bibliographystyle{apsrev4-1}
	\bibliography{1_bib}
	\newpage
	\onecolumngrid
	\appendix
	\section{Diamagnetic repulsion}\label{NN}
	    From Eq. (\ref{CoM}) one can immediately write down the diamagnetic repulsion given by the entire diamond,
	\begin{equation}
		\begin{aligned}
			F_\text{dia}^{\text{L,R}}&= \partial_X \frac{\chi_\rho M}{2 \mu_0}B^2(X)|_{X=X^\text{L,R}_\text{c}}\\
			&= \frac{\chi_\rho M}{ \mu_0}(B_0\eta_\text{b}+\eta^2_\text{b}X^\text{L,R}_\text{c})
		\end{aligned}\label{N1}
	\end{equation}
The diamagnetic force here is a macroscopic quantity, representing the repulsion of the entire diamond to the magnetic field, which effectively located at CoM with equilibrium position $X_\text{c}$. However not like the splitting force which acts on the central particle, this does not mean that the repulsion exhibited by particles in the middle of the atomic chain is equal to Eq. (\ref{N1}). 
	
	Here, we consider that each atom in the atomic chain is magnetized with the potential energy $U(X_i)$.  We will not delve into the specific form of this $U(X_i)$, but in general, it is the product of the magnetization of a single carbon atom and the magnetic field strength experienced by that atom (relevant discussions can be found in Chapter 4 of \cite{kittel2018introduction}). Therefore, each atom will contribute a different diamagnetic force, and these microscopic diamagnetic forces will vary in magnitude due to the different equilibrium positions of the atoms,
		\begin{equation}
		\begin{aligned}
			F_{i,\text{dia}}^{\text{L,R}}&= -\partial_X U(X)|_{X=X^\text{L,R}_i}
		\end{aligned}\label{N2}
	\end{equation}
	The summation over index $i$ of these microscopic diamagnetic forces must equal to Eq. (\ref{N1}), and thus we have the relation Eq. (\ref{Fdia}).

	\section{Initial thermal equilibrium mixed state}\label{appen2}
	Here, we provide supplementary information about characteristic length and characteristic velocity for phonons in (\ref{12}). Consider that before any interaction between fields, the quantum oscillator system for mode $q$ is, 
	\begin{equation}
		\begin{aligned}
			H_\text{free}=\frac{1}{2}(\dot{u}_q^2+\omega_q^2 u_q^2),
		\end{aligned}
	\end{equation}
	the eigen-energy of this simple Hamiltonian is obvious $E_n=\hbar \omega_q (n+\frac{1}{2})$ where $n$ is the phonon number in mode $q$.
	In this pure state, there is no ensemble to work with. Now, let us consider that initially, there are various phonon modes live in the diamond where they have reached a thermal equilibrium state with partition function,
	\begin{equation}
		\begin{aligned}
			Z&=\sum_n e^{-\beta E_n}\\
			&=e^{-\frac{\beta}{2} \hbar \omega_q} \sum_{n=0}^{\infty} e^{-n \beta \hbar \omega_q}\\
			&=\frac{1}{2 \sinh \left(\frac{\beta}{2} \hbar \omega_q\right)}
		\end{aligned}\label{trick}
	\end{equation}
	where $\beta=1/k_\text{B}T$. 
	The calculation from the second to the third line in Eq. (\ref{trick}) can be found in, for example, \cite{kittel2018introduction}. The average energy is, thus, 
	\begin{equation}
		\langle E \rangle_q=-\frac{\partial \ln Z}{\partial \beta}=\frac{\hbar \omega_q}{2} \operatorname{coth}\left(\frac{\beta}{2} \hbar \omega_q\right)
	\end{equation}
	Note that the equipartition theorem tells us that the collection of phonon's potential energies and kinetic energies are the same, such that we have, 
	\begin{equation}
		\left\langle\frac{\dot{u}_q^2}{2 }\right\rangle=\left\langle\frac{\omega_q^2 u_q^2}{2}\right\rangle=\frac{\langle E \rangle_q}{2}=\frac{\hbar \omega}{4} \operatorname{coth} \frac{\hbar \omega}{2 k_{\mathrm{B}} T} .
	\end{equation}
	Therefore, the square of characteristic width and velocity becomes, 
	\begin{equation}
		\sigma^2_{\dot{u}_q}=\left\langle\dot{u}_q^2\right\rangle=\frac{\hbar \omega}{2} \operatorname{coth} \frac{\hbar \omega_q}{2k_\text{B} T},~~~	\sigma_{u_q}^2=\left\langle u_q^2\right\rangle= \sigma_{u_q}^2=\frac{\hbar}{2 \omega_q} \operatorname{coth} \frac{\hbar \omega_q}{2k_{\mathrm{B}} T}
	\end{equation}
	\section{Overlap and Wigner representation}
	Here, we supplement some details about Eq. (\ref{hamiltonianq3}). We can represent the time evolution of the phonon wave function in the interaction picture for a specific mode $q$ as,
	
	\begin{equation}
		\begin{aligned}
			|\psi^\text{L,R}_{\text{I},q}(t)\rangle= \exp\left(-\frac{\text{i}}{\hbar}\int_0^t \text{d}t^\prime V^\text{L,R}_{q}(t^\prime)\right)|\psi^\text{L,R}_{q}(0)\rangle
		\end{aligned}\label{A1}
	\end{equation}
	We have set $|\psi_{\text{I},q}(0)\rangle= |\psi_q(0)\rangle$ where $|\psi_q(0)\rangle$ is a stationary state in the Schrödinger picture. We have also ignored a global phase factor which does not contributes to the contrast. One can express the interaction potential in terms of phonon mode amplitude and momentum as,
	\begin{equation}
		\begin{aligned}
			V^\text{L,R}_q(t)&= -\sum_i F_i^\text{L,R} Q_q^i\sqrt{\frac{\hbar}{2M \omega_q}}(\hat{a}_{-q}^\dagger \text{e}^{\text{i}\omega_qt}+\hat{a}_q \text{e}^{-\text{i}\omega_qt})\\
			&= -\sum_i \left[\hat{u}_q\frac{F_i^\text{L,R}}{\sqrt{M}}Q_q^i \cos(\omega_qt)+ \hat{\dot{u}}_q \frac{F_i^\text{L,R}}{\sqrt{M}}Q_q^i \frac{\sin(\omega_qt)}{\omega_q}\right]\\
			&=- \hat{u}_q  \cos(\omega_qt) f^\text{L,R}_q- \hat{\dot{u}}_q  \frac{\sin(\omega_qt)}{\omega_q} f^\text{L,R}_q\\
		\end{aligned}
	\end{equation}
	Where $f_q^\text{L,R}$ has already been defined in Eq. (\ref{13}) and the relation between creation, annihilation and phonon amplitude can be found in (\ref{Free1}) and (\ref{Free2}).
	The overlap of the wavefunctions of phonons on both sides of the interferometer arm can be expressed as:
	\begin{equation}
		\begin{aligned}
			&	\langle\psi^\text{L}(\Delta t)|\psi^\text{R}(\Delta t)\rangle\\
			&=\langle\psi_\text{I}(0)|\exp\left\{\frac{\text{i}}{\hbar}\int_0^{\Delta t} \text{d}t^\prime \left[V^\text{L}_{q}(t^\prime)-V^\text{R}_{q}(t^\prime)\right]\right\}|\psi_\text{I}(0)\rangle\\
			&= \langle\psi_\text{I}(0)|\exp\left\{\frac{\text{i}}{\hbar}\int_0^{\Delta t} \text{d}t^\prime \left[\hat{u}_q  \cos(\omega_qt) \Delta f_q+ \hat{\dot{u}}_q  \frac{\sin(\omega_qt)}{\omega_q} \Delta f_q    \right]\right\}|\psi_\text{I}(0)\rangle\\
		\end{aligned}\label{b3}
	\end{equation}
	The integral terms in the third line of Eq. (\ref {hamiltonianq3}) precisely represent the differences in phonon amplitudes and momentum given in Eq. (\ref{difference}). 
	Here, one can observe that if the force acting on the atomic chain is coupled with spin and magnetic field, then $V_\text{I}^\text{L}=- V_\text{I}^\text{R}$. For diamagnetic interaction, since the diamagnetic force points in the same direction, the situation is different.

	We use the Wigner representation to calculate the average of (\ref{hamiltonianq3}). under the Wigner representation, the average of an operator $\hat{A}(x)$ can be calculated as, 
	\begin{equation}
		\begin{aligned}
			\langle \hat{A} \rangle = \int\int W(x,p)\tilde{A}(x,p) \text{d}x \text{d}p
		\end{aligned}\label{average}
	\end{equation}
	where $W(x,p)$ is the Wigner function, it provides a quasi-distribution of quantum state simultaneously in position $x$ and momentum $p$ space. Usually, in the Wigner representation, one typically needs to utilize the Weyl transform to express the operator $\hat{A}$ as a function $\tilde{A}(x, p)$. In this context, the exponential operator in (\ref{b3}) is already a function of phonon normal coordinates and their conjugates. We first consider a pure state of phonons in mode $q$, and its ground state wavefunction can be expressed as follows,
	
	\begin{equation}
		\psi\left(u_q\right)=\left(\frac{\omega_q}{\pi \hbar}\right)^{1 / 4} e^{-\omega_q u_q^2 / 2 \hbar}.
	\end{equation}
	The Wigner function is defined as,
	\begin{equation}
		\begin{aligned}
			W\left(u_q, \dot{u}_q\right) & =\frac{1}{2 \pi \hbar} \int_{-\infty}^{\infty} e^{\text{i} \dot{u}_q y} \psi\left(u_q+\frac{y}{2}\right) \psi^*\left(u_q-\frac{y}{2}\right) \mathrm{d} y \\
			& =\frac{1}{\pi \hbar} \exp \left[-\left(\frac{u_q^2}{2\left(\frac{\hbar}{2 \omega_q}\right)}+\frac{\dot{u}_q^2}{2\left(\frac{\hbar \omega_q}{2}\right)}\right)\right]
		\end{aligned}\label{Wigner1}
	\end{equation}
	where $\frac{\hbar}{2 \omega_q}$ and $\frac{\hbar \omega_q}{2}$ relate to characteristic width and velocity, respectively. In the thermal equilibrium mixed state, one needs to replace,
	\begin{equation}
		\begin{aligned}
			\frac{\hbar \omega}{2}\rightarrow	\frac{\hbar \omega}{2} \operatorname{coth} \frac{\hbar \omega_q}{2k_\text{B} T},~~~~~~~\frac{\hbar}{2 \omega_q}\rightarrow\frac{\hbar}{2 \omega_q} \operatorname{coth} \frac{\hbar \omega_q}{2k_{\mathrm{B}} T}.
		\end{aligned}
	\end{equation}
	\section{Contrast loss due to intrinsic dipole-phonon coupling}\label{dipole1}
	
	In this section, we present the contrast loss due to dipole-phonon coupling. It's worth noting that the dipole-phonon coupling we consider does not involve energy level transitions of the dipole, such as in the Jaynes-Cummings model. We only consider the overlap of the positions and momenta of the phonons on both sides of the interferometer arm, that is, the motion state of the phonons.

	Here, let's first consider the case of an intrinsic dipole. In diamond materials embedded with NV centres, due to the NV centre itself carrying a negative charge, the inevitable occurrence of compensatory positive charges arises when it replaces a carbon atom in the material. 
	As a result, a dipole moment $d_0$ arises, and its specific value depends on the distance from the compensatory positive charges to the NV centre. This value is approximately one Debye.
	Here, we aim to study the effect of the dipole on lattice vibrations through its interaction with an external electric field. Therefore, we assume that this intrinsic dipole overlaps with the NV centre, meaning that the compensatory positive charges and the NV centre are located in the same lattice site.
	
	Same to the analysis in the main text, let us denote the position of a single intrinsic dipole as $X_c(t)+x_c$, where $X_c(t)$ represents the equilibrium position while $x_c$ represent dipole's vibration. The dipole-phonon interaction Hamiltonian is again mediated by an extra electric field $E(X_c)$ as.
	\begin{equation}
		\begin{aligned}
			H_\text{Intrinsic}=-d_0 E(X_c)\\
		\end{aligned}
	\end{equation}
	Recall that the diamond has its trajectories as illustrated in Fig. \ref{figure2}, the $X_c$ can be expressed explicitly as (\ref{trajectories}) and $E(X_c)$ represents the electric field experienced by the dipole. 
	
	Again, due to the dipole being in a spatial superposition state (considering the diamond entering the SG apparatus), if there is a difference in the derivatives of $E(X_c)$ on both sides of the interferometer arm, there will be differences in the amplitudes and momentum of the phonons, as illustrated in (\ref{difference}). Obviously, if $E(X_c)$ is linear, the coupling between the dipole and the electric field will not cause such differences.
	
	\section{Contrast loss due to induced dipole-phonon coupling}\label{dipole2}
	This section presents an analysis of the induced dipole-phonon coupling on phonon contrast loss. Similarly, we only consider the motion state of the phonons and do not consider the energy level issues of the dipole. We consider that each particle $i^\text{th}$ in the atomic chain experiences an external linear electric field,
	\begin{equation}
		\begin{aligned}
			E(X_i)= E_0+\eta_\text{e}X_i(t)
		\end{aligned}
	\end{equation}
	where the constant electric gradient is labeled as $\eta_\text{e}$. Again, particles move parallel to the diamond's trajectories, and  their equilibrium positions $X_i$ become functions of time. Due to the fact that the diamond is a dielectric material when subjected to the external electric field, an induced dipole will be generated in each atom of the lattice chain as,
	\begin{equation}
		\begin{aligned}
			d_i(X_i)=\frac{\alpha}{\epsilon_r} E(X_i)
		\end{aligned}
	\end{equation}
	where $\epsilon_r\approx 5.7$ is the relative dielectric constant and the polarizability $\alpha$ is given by the Claussius-Mossotti relation, 
	\begin{equation}
		\frac{\alpha N}{3 \epsilon_0 V}=\frac{\epsilon_{\mathrm{r}}-1}{\epsilon_{\mathrm{r}}+2} .
	\end{equation}
	where $V$ is the volume of the diamond. Therefore, the interaction Hamiltonian with the external electric field can be expressed as,
	
	\begin{equation}
		\begin{aligned}
			H_\text{Induced}&=-\sum_i d_i(X_i) E(X_i)\\
			&= -\sum_i \frac{\alpha}{\epsilon_r} E^2(X_i)
		\end{aligned}
	\end{equation}
	
	The force exerted on $i^\text{th}$ particle can be written straightforwardly as,
	\begin{equation}
		\begin{aligned}
			F^\text{dp}_i&= \frac{2\alpha}{\epsilon_r}E_0\eta_\text{e} +\frac{2\alpha}{\epsilon_r}\eta^2_\text{e}X_i \label{D1}
		\end{aligned}
	\end{equation}	
	The force acting on each polarized particle can be divided into two parts. The first term is the same on both sides of the interferometer arm, while the second term differs due to the spatial superposition state of the diamond. It should be noted that $F_i$ here represents the force experienced by each polarized atom, and depending on their equilibrium positions, the force on each atom is different. However, since in this model, $X_{i+1}-X_i=\text{const.}$ where const. represents the distance between two atoms, therefore we have $X_i^\text{L}-X_i^\text{R}=\Delta X_c$.
	Substituting (\ref{D1}) into (\ref{Contrastnew}) and (\ref{TransformFunction}), one can obtain the contrast loss caused by induced dipole-phonon interaction as, 
	\begin{equation}
		\begin{aligned}
			C_\text{dp} & =\prod_q \exp \left[-\frac{\operatorname{coth} \frac{1}{2} \frac{h \omega_q}{k_B T}}{ \hbar \omega_q} \times\left|\frac{3 V \Delta X_{\mathrm{m}} \eta_{\mathrm{e}}^2}{\tau_{\mathrm{a}}^2 \omega_q^3}\frac{\epsilon_0\left(\epsilon_{\mathrm{r}}-1\right)}{\epsilon_{\mathrm{r}}\left(\epsilon_{\mathrm{r}}+2\right)} \times \Gamma\left(\omega_q\right)\right|^2\right] \\
		\end{aligned}\label{D6}
	\end{equation}
	where, $\Gamma\left(\omega_q\right)$ is related to the Fourier transform given in (\ref{transform}). 
	Here, one can estimate that the contrast loss induced by the induced dipole is very weak. Taking the diamond masses of interest ($10^{-20}$ to $10^{-14}$ kg) as an example, their phonon frequency ranges from tens of gigahertz to several terahertz. In this case, the first term in (\ref{D6}) is approximately on the magnitude of $10^{-24}$. 
	For the absolute value term, even if one chooses the maximum value of $ \Gamma(\omega_q)$ and specifies a splitting protocol that achieves a maximum splitting of around $100 \mu$m within one second, the order of magnitude of this term only reaches about $10^{-130}\times \eta_\text{e}^4$. Consider a case where a single charge is held at a distance of 500$\mu$m away from the diamond, in this case the electric gradient reaches only to $30$V/m$^2$. Even if one assumes that the distance between the free charge held from the diamond reaches $100$nm, the electric field gradient at this point is approximately $\sim 10^{10}$V/m$^2$, which is not sufficient to increase the contrast caused by dipole-electric field-phonon coupling becomes negligible.

\end{document}